\documentstyle[aps,twocolumn,graphics,epsf,amsmath,amssymb,eqsecnum,floats]{revtex}
\begin{document}

\title{ 
 Spin-1/2 frustrated antiferromagnet on a
 spatially anisotopic square lattice:
 contribution of exact diagonalizations.
}

\author{P. Sindzingre}
\address{Laboratoire de Physique Th{\'e}orique des Liquides-UMR 7600
of CNRS,  Universit{\'e}  Pierre  et  Marie  Curie,  case 121,   4 place
Jussieu, 75252 Paris Cedex, France\\ E-mail:phsi@lptl.jussieu.fr}
\maketitle
\bibliographystyle{prsty}
(\today)\\
\begin{abstract}
The phase diagram of a spin-1/2 $J-J'-J_2$ model 
is investigated by means of exact diagonalizations on finite samples.
This  model is a generalization of the $J_1-J_2$ model
on the square lattice with two different nearest-neighbor couplings $J,J'$
and may be also viewed as an array of coupled Heisenberg chains.
The results suggest that the resonnating valence bond state
predicted by Nersesyan and Tsvelik [Phys. Rev. B {\bf 67}, 024422 (2003)]
for $J_2=0.5J' \ll J$ is realized and extends beyond the limit
of small interchain coupling along a curve nearly coincident
with the line where the energy per spin is maximum.
This line is likely bordered on both sides by 
a columnar dimer long range order.
This columnar order could extends for $J'\rightarrow J$ 
which correspond to the $J_1-J_2$ model.
\end{abstract}
PACS numbers: 75.10.Jm; 75.50.Ee; 75.40.-s

\section{INTRODUCTION}
The interplay between frustration and quantum fluctuations 
in antiferromagnetic (AF) spin-1/2 systems in two dimensions (D) 
has attracted much experimental and theoretical attention 
in past years.
In particular, many studies have been devoted to identify
the possible phases which may appear at $T=0$ after the destabilization
of a collinear AF N\'eel phase by quantum fluctuations when 
increasing frustration in $SU(2)$ invariant spin models.
Large-$N$ approaches predict then a valence bond crystal (VBC) phase,
with long range order (LRO) of singlet entities 
(dimers, plaquettes of resonnating dimers...) 
and thus breaking translational symmetry and/or others 
space symmetries,
if the transition out of the N\'eel phase is continuous~\cite{rs90,rs91,sp01}.
Numerical results obtained from exact diagonalization (ED) studies
support this prediction for several models:
the $J_1-J_2-J_3$ model on the honeycomb lattice~\cite{fsl01} 
and some models where the spins sit on a square lattice such as
the crossed chain model~\cite{sfl02,fmsl01} and
the quadrumerized Shastry-Sutherland model~\cite{lws02}.

\begin{figure} [h!]
\hspace*{0.0cm}
\resizebox{7.0cm}{!}{\includegraphics{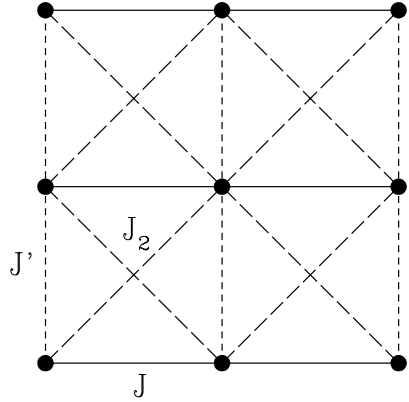}}\\

        \caption[99]{  Exchange interaction pattern in
the $J-J'-J_2$ model.
The  spins sit at the vertices shown by bullets,
the coupling constant is $J$ between nearest-neighbor pairs
on hozirontal bonds (full lines), $J'$ on vertical bonds
(short dashed lines)
and $J_2$ on the diagonal bonds (long dashed lines).
The $J_1-J_2$ model corresponds to $J'=J=J_1$.
        }  \label{fig_exchanges}
\end{figure}

\begin{figure}  [h]
\vspace{-4.3cm}
\hspace*{0.1cm}
\resizebox{8.0cm}{!}{\includegraphics{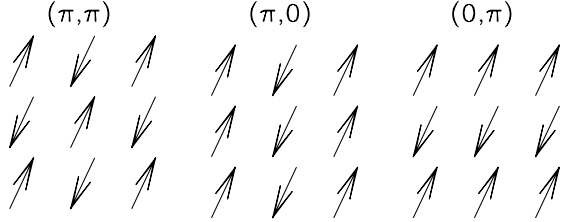}}\\
        \caption[99]{ Spins orders in the classical
$(\pi,\pi)$, $(\pi,0)$ and $(0,\pi)$ N\'eel states.
        }  \label{types_neel}
\end{figure}

\begin{figure}  [h]
\vspace{-4.3cm}
\hspace*{0.1cm}
\resizebox{8.0cm}{!}{\includegraphics{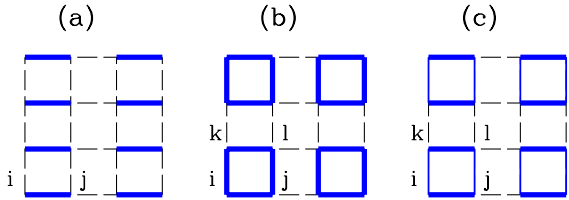}}\\
        \caption[99]{ 
Patterns of VBC orders proposed for $J_1-J_2$ model:
(a) columnar dimer,
(b) plaquette,  
(c) columnar  dimer with plaquette type modulation.
Full lines  connect spins bound into singlet: two spins in (a),
four spins in (b) and (c).
In (b) the strengh of spin-spin correlations on vertical and horizontal links
of a plaquette $i,j,k,l$ are the same but differs in (c).
There are four degenerate ground-states for orders (a), (b)
and eight for the order (c) in the $J_1-J_2$ model.
These are obtained by one step translations in the horizontal
and vertical directions and/or a $\pi/2$ rotation around 
a lattice site.
        }  \label{types_vbc}
\end{figure}

Yet, for the much studied $J_1-J_2$ model on the square lattice 
with a next-nearest-neighbor frustrating coupling $J_2$ 
in addition to the nearest-neighbor coupling $J_1$
(see Fig.~\ref{fig_exchanges} where the $J_1-J_2$ model corresponds to 
$J'=J=J_1$), the validity of this scenario remain 
unclear~\cite{sz92,szp96,swho99,kosw00,co00,cls00,sow01,cbps01,sow02,cbps03}.
Its classical ground-state has $(\pi,\pi)$ N\'eel LRO 
for $J_2 <0.5 J_1$
and is the degenerate manifold corresponding to two decoupled 
N\'eel states on interpenetrating square lattices for $J_2 >0.5 J_1$.
There is a large consensus that $(\pi,\pi)$ N\'eel LRO
(see Fig.~\ref{types_neel}) survives to quantum fluctuations 
for $ J_2 \alt 0.4 J_1$ and quantum fluctuations select
either a $(\pi,0)$ or a $(0,\pi)$ N\'eel state  in the manifold
of classical ground-states for $J_2 \gtrsim 0.65 J_1$,
whereas, in the intermediate range $0.4J_1\alt J_2 \alt 0.65J_1$ 
between these two gapless states,
there is a phase without N\'eel LRO gapped to spin excitations.
The transition out of the $(\pi,\pi)$ N\'eel phase is  second order
while the transition out of the $(\pi,0)$ N\'eel phase 
has been predicted to be of first order~\cite{swho99}
but may be close to second order~\cite{sow01}.

Whether this intermediate phase has VBC LRO or is 
some resonnating valence bond (RVB) phase
which do not break any space symmetries, i.e. a spin-liquid, is debated.
Earliest investigations, from ED calculations on samples up to $N=36$ spins,
have shown enhanced columnar dimer-dimer correlations
in this intermediate region, but
the finite-size scaling of the order parameter for the columnar VBC LRO
of dimers, shown in Fig.~\ref{types_vbc}(a),
were not conclusive~\cite{sz92,szp96}.
Dimer series expansions, however, were in favor of VBC LRO:
earliest studies favored columnar LRO~\cite{swho99,kosw00} while
latter calculations
have lead to propose the occurence of two different VBC phases: 
the columnar dimer phase for $J_2/J_1 \alt 0.5$ 
and a columnar dimer phase with plaquette type modulation,  
as shown in Fig.~\ref{types_vbc}(c), for $0.5 \gtrsim J_2/J_1$,
separated by a second order transition~\cite{sow01}.
On the other hand,
studies based on the Green function Monte Carlo method,
for systems of sizes larger than those
which may be investigated by ED calculations,
first favored either a plaquette LRO~\cite{co00},
as shown in Fig.~\ref{types_vbc}(b),
or a columnar dimer LRO with plaquette type modulation~\cite{cls00},
similar to one of Ref~\cite{sow01},
for $J_2/J_1=0.5$.
Besides, it was concluded in Ref~\cite{co00} that the
finite size scaling of the susceptibility to columnar dimer LRO
from both Monte Carlo and ED calculations
would exclude columnar dimer LRO at $J_2=0.5J_1$.
However, latter calculations~\cite{cbps01}, by the same group which favored 
the plaquette LRO~\cite{co00}, have lead to conclude in favor
of the occurence of a spin-liquid phase
for $0.4J_1\alt J_2\alt 0.5J_1$ whereas it was pointed out
that the ocurrence of the plaquette LRO will be 
unlikely in view of the spatial symmetries of the low lying singlets
in the ED spectra.
In their latest study these authors suggest either a spin-liquid
or, possibly, the columnar state but with a weak order parameter\cite{cbps03}
Yet, as the Monte Carlo calculations and the series expansions
start from some trial state which may bias the results
and involve approximations 
which are not fully under control,
while ED calculations which were performed  on samples having different
lattice symmetries and display irregular finite size scaling,
the nature of the ground-state in the intermediate region
remain an open question.

In this paper, motivated by a recent paper of 
Nersesyan and Tsvelik~\cite{nersesyan-tsvelik-03},
refered as (NT) in the following,
we attempt to shed light on this issue from ED calculations
on a spatially anisotropic version of the $J_1-J_2$ model 
with  two different  nearest neighbor couplings. 
The Hamiltonian of this $J-J'-J_2$ model 
(named the confederate flag model by NT)
reads:
\begin{eqnarray}
{\cal H} = & \sum_{m}^{M}   \sum_{l}^{L} [
                 \,J\,{\bf S}_{l,m}.{\bf S}_{l+1,m} \,+\,
                 \,J'\,{\bf S}_{l,m}.{\bf S}_{l,m+1} \,+\, \nonumber \\
               & \,J_2\, ( {\bf S}_{l,m}.{\bf S}_{l+1,m+1} \,+\,
                           {\bf S}_{l,m}.{\bf S}_{l+1,m-1}) \,]
\label{eq-Heis}
\end{eqnarray}
which may also be viewed as an array of $M$ spin chains of lenght $L$,
if $J'<J$.
The exchange $J$ (see Fig.~\ref{fig_exchanges})
couples first-neighbor (horizontal) pairs of spins
along the chains,
exchange $J'$ first-neighbor spins in the (vertical) 
transverse direction 
and exchange $J_2$ second-neighbor pairs on the diagonals of the 
square plaquettes.
All exchanges are assumed positive, describing AF couplings.
For $J'=J$ one recovers the $J_1-J_2$ model.
Mostly interested with the limit $M,L\rightarrow\infty$,
we shall also focus on the case $J'\le J$.

The classical ground-state of the $J-J'-J_2$ model is rather similar to the
one of the $J_1-J_2$ model.
When $J'<J$, one has $(\pi,\pi)$ N\'eel LRO if $J_2<0.5 J'$,
$(\pi,0)$ N\'eel LRO if $J_2>0.5 J'$ 
and the (horizontal) $J$ chains are decoupled 
if $J_2=0.5 J'$.
For $J'>J$, one has $(\pi,\pi)$ N\'eel LRO if $J_2<0.5 J$,
$(0,\pi)$ N\'eel LRO if $J_2>0.5 J$ 
whereas the (vertical) $J'$ chains are decoupled
if $J_2=0.5 J$.

This Hamiltonian has been much studied in past years for
a finite number of chains, the so-called $M$-leg ladders,
generally in the case of open boundary conditions in the
transverse direction 
(for $M>2$ since if $M=2$ open and periodic boundary conditions
are equivalent).

Numerical studies~\cite{wns94,gbw96,fat96} 
based on the density matrix renormalization group (DMRG) method 
or Monte Carlo approaches have shown that
the $M$-leg unfrustrated ladders ($J_2=0$) display
a behavior analogeous to a $M/2$-spin chain~\cite{haldane_83}, 
being fully gapped if $M$ is even and gapless if $M$ is odd.
So $(\pi,\pi)$ N\'eel LRO only occurs in the limit $M\rightarrow\infty$.

\begin{figure}  [h]
        \begin{center}
        \resizebox{7.5cm}{6.5cm}{\includegraphics{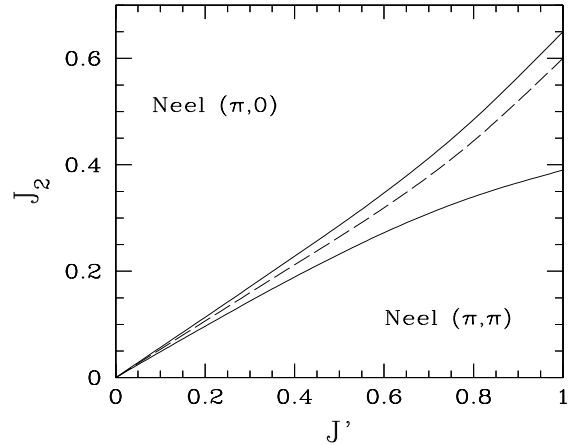}}
        \end{center}
        \caption[99]{  Proposed  phase diagram of the $J-J'-J_2$ model
 for $L\rightarrow\infty$ and $M\rightarrow\infty$  ($J=1$).
 The solid lines indicate the approximate location of the
 boundaries of the $(\pi,\pi)$ and $(\pi,0)$ N\'eel phases,
 noted $J_{2}^{c1}(J')$ and $J_{2}^{c2}(J')$ respectivelly.
 The dashed line in the intermediate region between the N\'eel phases
 is the approximate location of the maxima $J_{2}^{m}(J')$
 of the energy per spin $e_0$
 along which the present ED results suggest
 that the RVB state predicted by NT
 may be realized up to large values of $J'$.
 In the intermediate region, outside the curve $J_{2}^{m}$,
 the ground-state is predicted to display columnar dimer order.
 The N\'eel phases only appear for both $L,M\rightarrow\infty$.
 For finite $M$ and $L\rightarrow\infty$ one has gapped phases.
 The position of the curve $J_{2}^{m}(J')$ is quasi-independent
 of the number $M$ of chains down to  $M=2$, which correspond to 
 the $2$-leg ladder.
        }    \label{phase_diag}
\end{figure}

The effect of frustration has been especially investigated
theoretically and numerically for the $2$-leg ladder
(see Ref~\cite{wko98,wang98,hhr01,aen00} and references therein).
These studies have concluded that
one has two phases separated by a transition line which has been found to
coincide with the line of maximum frustration  
where the energy per spin reaches its maximum,
noted $J^{m}_2(J')$ in the following for all values of the number of chains.
This line starts as $J^{m}_2(J')\sim 0.5 J'$ for $J'\rightarrow 0$
and bend slightly with increasing $J'$ so 
that $J^{m}_2(J')\sim 0.6$ at $J'=1$~\cite{wko98,wang98}
(as the dashed line in Fig.~\ref{phase_diag}).
If $J_2 > J^{m}_2(J')$ one has the 'Haldane phase',
where the two spins on a rung tend to be in a triplet state,
so named as it contains the point $J_2=J, J'=0$ whose
low-energy spectrum is similar to that of a spin-1
Heisenberg chain.
If $J_2 < J^{m}_2(J')$  on has the so called 'singlet phase',
so called as it contains the case $J'\gg J$ with $J_2=0$,
where the ground-state consists of singlets on each rung.
As shown by White~\cite{white96} the 'singlet phase' 
is also an Haldane phase with diagonally situated 
next-nearest neigbor spins coupling to form an effective $S=1$.
Both phases have a topological order,
breaking a hidden $Z_2\times Z_2$ symmetry,
which can be measured  by non-local 'string order' parameters,
similar to the string order parameter of the $S=1$ Heisenberg chain.
The two phases are fully gapped like the $S=1$ chain~\cite{haldane_83},
and have a non degenerate singlet ground-state.
These ground-states become degenerate on the transition line $J^{m}_2(J')$.
The nature of the transition has been debated.
In a DMRG study ~\cite{wang98} 
it was predicted that the transition line would be gapless
for $J_2$ lower than $\approx 0.287 J$. 
But more recent DMRG calculations~\cite{hhr01} 
have concluded that a gap subsits
between the two-fold degenerate ground-state and the excited states 
for all $J_2>0$.
So the transition is 1th order on the whole line $J^{m}_2(J')$ 
in agreement with bosonization results in the limit
of vanishing interchain coupling~\cite{aen00}.
The elementary excitations then 
consist of gapped deconfined spin-1/2 spinons
which may be viewed as kinks interpolating between the two 
ground-states~\cite{aen00}.
Note that these results imply that the classical behavior of independent chains
on the line $J_2=0.5 J'$ is then destabilized by quantum fluctuations.
The single Heisenberg chain is gapless. 
The spectrum of two independent chains too.
The existence of a finite gap excludes such a behavior.

A DMRG study of the three-leg frustrated ladder with open boundary
conditions in the transverse direction~\cite{wzc02}
has shown that this system 
exhibit two phases separated by a transition line
also coincident with the line of maximum frustration $J^{m}_2(J')$ and
which is a curve quasi identical to the one found for the 2-leg ladders.
The phases are analogeous to those of the 2-leg ladders.
The small $J_2$ phase, which is the phase of the unfrustrated ladder,
show a tendency of the three spins on a
vertical line to pair in a state of minimum spin
whereas they tend to pair in a state of maximum spin in the 
large $J_2$ phase, which is equivalent to the spin-3/2 chain.
On the transition line a bosonization study at weak interchain 
coupling~\cite{aln98} has predicted that the ground-state
is a chiral spin-liquid.

Recently NT, using an approach based on the bosonization method,
investigated the crossover from finite $M$ to $M\rightarrow\infty$ 
in the $J-J'-J_2$ model  on the line $J_2=0.5J'$,
in the limit of small interchain coupling ($J',J_2\ll J$)
and predicted  that the classical behavior of independent chains
is unstable to a special RVB (spin-liquid) state 
which may be a realization of the chiral $\pi$-flux RVB 
state ~\cite{nersesyan-tsvelik-03,smirnov-tsvelik-03,bhaseen-tsvelik-03}.
Its ground-state has a $2^{M-1}$ degeneracy,
for $M$ even and transverse periodic boundary conditions,
associated to the breaking of a local $Z_2$ symmetry, 
present in the bosonised version of the model,
corresponding  to the invariance under 
independent translations by one lattice spacing along individual chains.
Each ground-state may be caracterized by different values of 
a set of non-local order parameters.
These non-local order parameters correlate spins on two adjacent chains 
whereas different pairs of chains are uncorrelated. 
The elementary excitations consist of deconfined spin-1/2 spinons
interpolating between different ground-states.
These excitations are either gapless, as predicted for some values of the
number of chains, such as $M=4,6,12$, or gapped, as for 
the $2$-leg ladder ($M=2$).

Two questions then arise. 
First, does the state predicted by NT extends outside the range of 
small interchain coupling.
Second, what is the nature of the behavior in its vicinity,
in particular for the case of an infinite number of chains.
Below we report results of ED calculations 
of $N=M\times L$ spins for $N\le 36$ to shed light on
these questions and  more generally
the phase diagram of the $J-J'-J_2$ model.
We shall consider only the case of an even number of chains.
The ED calculations have been carried out on samples of $N=M\times L$ spins
in $M$ chains of lenghts $L$  with periodic boundary conditions both along
and perpendicular to the chains: samples of 4 chains with
$N=16,24,32,36$ spins and 6 chains with $N=24,36$ spins.
 Additional calculations were also performed for the $M=2$ ladder
 for $N\le 32$ in order to compare with the $M=4,6$ results.
Contrary to the samples considered in previous ED calculations
for the $J_1-J_2$ model,
all samples have translation vectors parallel to the basis vectors
of the square lattice which leads
to a regular evolution of the properties with increasing $L$ and $M$.

Examination of the ED spectra
indicates that the model will display two N\'eel phases,
with magnetic wave-vectors $(\pi,\pi)$ and $(\pi,0)$,
separated by a magnetically disordered intermediate region,
in the limit $N\rightarrow\infty$ with both
$L\rightarrow\infty$ and $M\rightarrow\infty$,
as shown in Fig.~\ref{phase_diag}.
At variance with a study based on the DRMG method~\cite{moukouri03},
which appeared very recently on a preprint server,
our results provide indications that the 
classical behavior of independent chain is destabilized
by quantum fluctations and the RVB state of NT may
occur along a curve coincident with the line of maximum frustration
$J^{m}_2(J')$, located in this intermediate region,
up to large interchain coupling.
They also indicate that the intermediate region of the $M\rightarrow \infty$
phase diagram (see Fig.~\ref{phase_diag}),
on both sides of this line, has columnar VBC LRO,
which may extend up to $J'\rightarrow J$ 
and could already appear for $M$ even $\ge 4$.

In Sec.II we describe the evolution of the energy per spin
vs $J_2$ and $J'$ and the location of the curve $J^{m}_2(J')$.
The properties of the model in the regions of $(\pi,\pi)$ N\'eel LRO,
$(\pi,0)$ N\'eel LRO and in the intermediate region
are presented in Sec. III, IV and V, respectively.
Sec. VI gives a summary of the our results and discuss briefly
the possible behavior of the model for ferromagnetic $J_2$ and $J'$.

\begin{figure} [h] 
        \begin{center}
        \resizebox{7cm}{7cm}{\includegraphics{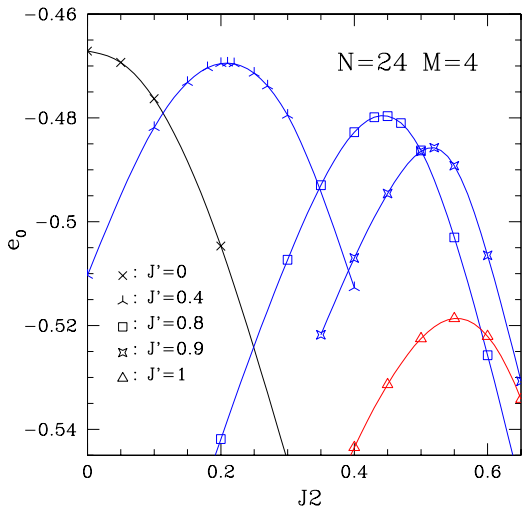}}
         \end{center}
        \caption[99]{ Ground-state energies per spin
$e_0=E_0/N$ vs $J_2$ for the $N=24$ sample with $M=4$ (4 chains).
The lines are guides to the eyes.
        }  \label{e0_n24_m4}
\end{figure}
\begin{figure} [h] 
        \begin{center}
        \resizebox{7cm}{7cm}{\includegraphics{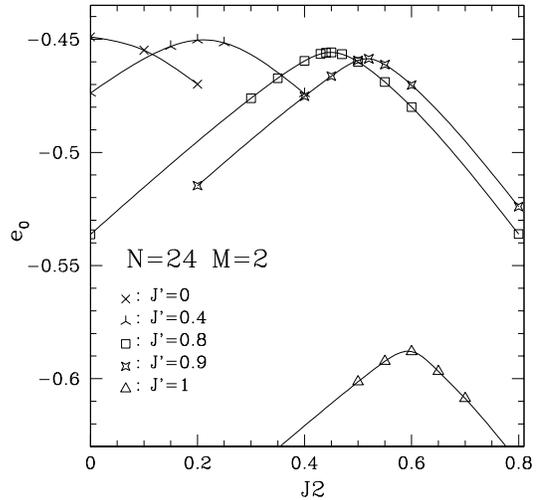}}
         \end{center}
        \caption[99]{ Same as Fig.~\ref{e0_n24_m4}
for the $N=24$ sample with $M=2$ ($2$-leg ladder).
        }  \label{e0_n24_m2}
\end{figure}

\section{Energy results}

\begin{figure} [h] 
        \begin{center}
        \resizebox{7cm}{7cm}{\includegraphics{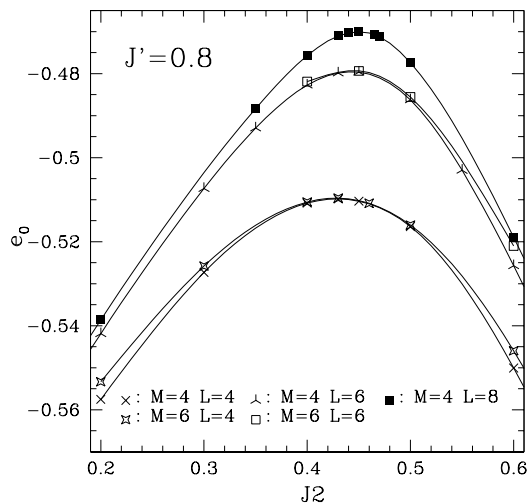}}
        \end{center}
        \caption[99]{Ground-state energies per spin
$e_0=E_0/N$ vs $J_2$ at $J'=0.8$ for different samples
of $M$ chains of lenghts $L$.
        }  \label{e_j2_0.8_vs_j}
\end{figure}

The ground-state energies per spin $e_0=E_0/N$ 
are ploted, as a function of $J_2$, for different values of $J'\le 1$
($J=1$ in the following)
for the $M=4$ chains, $N=24$, sample in Fig.~\ref{e0_n24_m4}.
For comparison,
the results for  the $N=24$  $2$-leg ladders are indicated 
in Fig.~\ref{e0_n24_m2}.
Fig.~\ref{e_j2_0.8_vs_j}, displays $e_0$ vs $J_2$ at $J'=0.8$
for the different $M=4$ and $M=6$ samples.
As shown in Fig.~\ref{e0_n24_m4} and Fig.~\ref{e_j2_0.8_vs_j},
the point of "maximum frustration" where $e_0$ reaches its maximum, 
$J_{2}^{m}(J')$ occurs for a value of $J_{2}$ slightly larger than 
$0.5 J'$, 
as found for the two and three-leg 
ladders~\cite{wko98,wang98,wzc02}.

As shown in  Fig.~\ref{e_j2_0.8_vs_j},
$J_{2}^{m}$ slightly increases with $L$, but is nearly independent
of the number $M$ of chains. 
This indicates that the  $L\rightarrow\infty$ curves $J_{2}^{m}(J')$
will be quasi-independent of $M$, as previously conjectured
from a comparison of the results 
for the two and three-leg ladders~\cite{wzc02}.
The maxima of $e_0$ (for a given sample) decrease with increasing $J'$
but remain rather close to the value for decoupled chains at $J'=0$
at least if $J'\le 0.9$.

In Fig.~\ref{e0_n24_m4} and Fig.~\ref{e0_n24_m2}
one sees, in the range $0.9<J'<1$ a drop of $e_0$ at $J_{2}^{m}$ 
from a value close to the one
for decoupled $J$ chains of lenghts $L$ to a value closer to the one
for independent $J'$ chains of lenghts $M$. This corresponds 
to the crossover to the situation at $J'\agt 1$ where the finite samples
$M \times L$ are best viewed as $L$ coupled chains of finite lenghts $M$.

\section{ $(\pi,\pi)$ N\'eel long range order }
\begin{figure} 
        \begin{center}
        \resizebox{7cm}{7cm}{\includegraphics{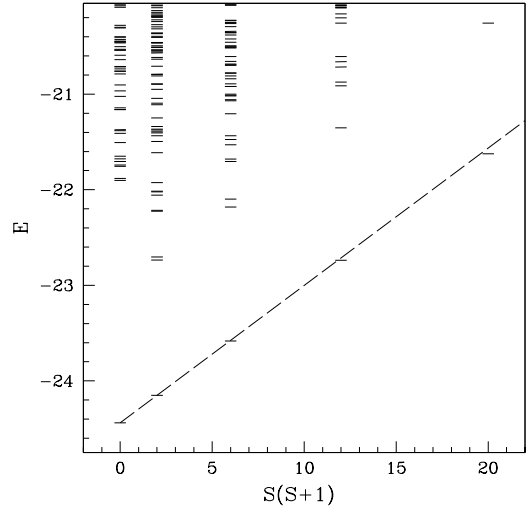}} \end{center}
        \caption[99]{ Spectrum 
at $J'=1$ and $J_2=0$ 
vs $S(S+1)$ for $N=36$.
The QDJS characteristic of collinear N\'eel LRO
at the bottom of each $S$ sector are well aligned (dashed line)
and clearly separated from the other lowest excitations (see text).
        }  \label{spec_sqr_heis36}
\end{figure}

Increasing $J_2$ from zero at given $J'$ ($\le 1$),
one finds a range of parameters, 
which extends to a value we note $J_{2}^{c1}(J')$, slightly smaller than 
$0.5J'$ (and thus $< J_{2}^{m}(J')$), 
where the spectra exhibit the features at finite size specific of
a system displaying collinear  $(\pi,\pi)$ N\'eel LRO
in the limit $N\rightarrow\infty$ with both
$L\rightarrow\infty$ and $M\rightarrow\infty$.
The N\'eel order breaks $SU(2)$ and lattice spacial symmetries.
Evidence of N\'eel LRO in the spectra is the presence of
lowest eigenlevels in each spin $S$ sector 
which form a set of states with energies $E(S)-  E_0 \sim S(S+1)/N$
for  $L\rightarrow\infty$ and $M\rightarrow\infty$,
the so-called quasi degenerated joint states (QDJS)~\cite{blp92,sfl02}
as shown in Fig.~\ref{spec_sqr_heis36}.
These QDJS are well separated, at finite size, 
from the others lowest excited states (magnons) 
and collapse on the ground-state faster with $N$ than the 
magnons, enabling the breaking of $SU(2)$ and lattice spacial symmetries.
For collinear LRO there is one QDJS for each $S$ value.
The QDJS specific of  $(\pi,\pi)$ N\'eel LRO consist of a state  belonging to
an irreducible representation (IR) with wave-vector
${\bf k}=0$ if $S$ is even or ${\bf k}=(\pi,\pi)$ if $S$ is odd, both
even in a spatial $R(\pi)$ rotation around a site and in the 
reflexion $\sigma$ with respect to a $J$ chain.

The approach of ED calculations does not allow an accurate
location of boundary of the N\'eel phase since it is limited
to small samples.
Nevertheless the extension of the N\'eel phase can be approximatively 
estimated from ED calculations 
by monitoring the range of values of $J_2$ for
which the QDJS remain well defined in the spectrum~\cite{sfl02}.
This gives $J_{2}^{c1}(J'=1)\approx 0.4$ for the $J_1-J_2$ model,
$J_{2}^{c1}(J'=0.8) \approx  0.35$.
$J_{2}^{c1}(J')$ moves steadily closer to the line 
$J_2=0.5J'$ as $J'$ decreases.
Besides, since
quantum Monte Carlo calculations on the unfrustrated line of the phase diagram
($J_2=0$) have indicated that $(\pi,\pi)$ LRO appear as soon as $J'>0$
for $M\rightarrow\infty$~\cite{sandvik99},
N\'eel LRO extend down to the point 
$J'=0, J_2=0$.
Thus the curve $J_{2}^{c1}(J')$ is tangent to the line $J_2=0.5J'$
for $J'\rightarrow0$,
moving away from this line as $J'$ increases from zero,
and is located  below the line,
as shown in Fig.~\ref{phase_diag}.

\begin{figure} 
        \begin{center}
        \resizebox{7cm}{7cm}{\includegraphics{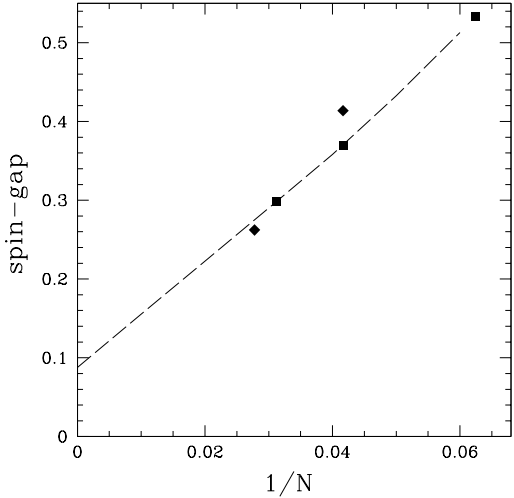}}
        \end{center}
        \caption[99]{ Spin-gaps
at $J'=0.8$ and $J_2=0$ vs $1/N$ for
for $N=16,24,32$ samples of $M=4$ chains (squares)
and the $N=24,36$ samples of $M=6$ chains (diamonds).
The dashed line is a spline fit to the $M=4$ values for $N \le 32$
followed by linear extrapolation for $N > 32$.
        }  \label{spingap_j2_0.8_j3_0.0}
\end{figure}

It is to be emphasized that N\'eel LRO appear only for both
$M\rightarrow\infty$ and $L\rightarrow\infty$.
There is no N\'eel LRO for a finite number $M$ of chains.
Indeed, as mentioned above, systems which consist of
a finite number $M$ of $S=1/2$ Heisenberg chains
are known to acquire a gap upon shifting a tranverse coupling $J'$
if $M$ is even, 
behaving analogeously to a spin $S=M/2$ chain,
so these systems are in a Haldane phase~\cite{wns94,gbw96,fat96,sandvik99}.

The typical evolution of the spin-gap $\Delta^{1}$ with $M$ and $L$
is illustrated in Fig.~\ref{spingap_j2_0.8_j3_0.0}
where values of the $\Delta^{1}$ are ploted vs $1/N$ 
for $N=16,24,32$ samples of $M=4$ chains and the $N=24,36$ samples
of $M=6$ chains at $J'=0.8$ and $J_2=0$.
The evolution for $M=4$ of $\Delta^{1}$ vs $1/N \sim 1/L$ 
shows a small but noticable upward curvature 
as found previously  for the 2-leg and 4-leg ladders 
(see Fig.1 of Ref.~\cite{wns94}).
As found in studies of the 2,4,6 legs ladders,
one may assume that the gap decreases monotonically as
$1/L$ decreases with a curvature that  remains positive
and vary only smoothly.
To estimate a lower bound on $\Delta^{1}(L\rightarrow\infty)$ we use
a linear extrapolation vs $1/L$, beyond  $N>32$ of a spline fit to
the $M=4$ three values of $\Delta^{1}$ 
(dashed line in Fig.~\ref{spingap_j2_0.8_j3_0.0}).
This leads to a lower bound of $\Delta^{1}$ for $L\rightarrow\infty$
$\sim 0.09$.
This lower bound
indicates that $\Delta^{1}(L\rightarrow\infty)$ remains finite.
Due to the neglect of the curvature beyond  $N>32$ the gap is
slightly underestimated.
The true value of $\Delta^{1}$ for $L\rightarrow\infty$ for $M=4$
is likely somewhat closer to the value ($\sim 0.14$) 
estimated for the 4-leg ladder with open boundary conditions
in the transverse direction~\cite{gbw96}.
For $M=6$, the present data, limited to the two values for $L=4,6$,
show that $\Delta^{1}(L\rightarrow\infty)$ decreases with increasing $M$,
and remain finite if $M=6$. 
Its value for the 6-leg ladder is  $\sim 0.04$~\cite{gbw96} 
if $J'=0.8$. 
The present data show that the gap is very small
but do not allow an accurate estimation of a gap of this
order of magnitude.

The evolution of the spin-gap with $N\rightarrow\infty$ for $M=L$ is
different than for fixed $M$ and $L\rightarrow\infty$.
For $M=L$, where $\Delta^{1}\rightarrow 0$ for $N\rightarrow\infty$,
the $N=16,36$ values in Fig.~\ref{spingap_j2_0.8_j3_0.0} indicates
an evolution of the spin-gap in the case $M=L$ that decreases 
with a negative curvature vs $1/N$ as usually found
in systems with N\'eel LRO~\cite{hn93}.

\section{ $(\pi,0)$ N\'eel long range order }
For $J_2$ larger than a value $J_{2}^{c2}$,
larger but very close to $J_{2}^{m}$,
one similarly recognizes in the spectra,
the features specific of collinear N\'eel LRO  with a wave vector 
$(\pi,0)$ or $(0,\pi)$ depending on value of $J'/J$.

If $J'=J=1$ and $M=L$ these two N\'eel orders are degenerate.
The QDJS then include two states for each value of $S$:
two with ${\bf k}=0$ if $S$ is even and
two with ${\bf k}=(\pi,0)$ and ${\bf k}=(0,\pi)$ if $S$ is odd,
all  even in $R(\pi)$ and $\sigma$.

The degeneracy is lifted if $J'\ne J$.
For $J'<J$ and in the thermodynamic limit $M=L\rightarrow\infty$ 
the $(\pi,0)$ LRO is favored whereas this would be  $(0,\pi)$ LRO for $J'>J$.
The QDJS of $(\pi,0)$ LRO are observed as the lowest states of the spectra
as soon as $J'<J$ for $M=L$.
They include only one state 
for each value of $S$: a state with ${\bf k}=0$ if $S$ is even 
and a state with ${\bf k}=(\pi,0)$ if $S$ is odd, 
both even in $R(\pi)$ and $\sigma$. 

Note, however, that
at finite size, for an aspect ratio $M/L<1$ of the samples,
quantum fluctuations, which favor an antiferromagnetic
allignement of the spins in the shortest direction, lead to
spectra caracteristic of $(0,\pi)$ LRO even for $J'<J$ as $J'\approx J$.
The transition between the two kind of spectra 
then occurs in the range of values of $J'$ where $e_0$ at $J^{m}_2(J')$
drops to a value similar to the one of $L$ decoupled chains
(see Fig.~\ref{e0_n24_m4}).
For $M=4$ and $L>M$, it is only for  $J' \alt 0.9$ 
that the QDJS of $(\pi,0)$ LRO
becomes lower than the QDJS of $(0,\pi)$ LRO.

The boundary of the $(\pi,0)$ N\'eel phase can be also estimated
in the same way as for the $(\pi,\pi)$ phase.
For the $J1-J2$ model it was found that $J_{2}^{c2}$ occurs at a value
slightly larger than $J_{2}^{m}\approx 0.6$, in the range 
$0.60 < J_{2}^{c2}(J'=1)< 0.70$. 
Examination of the spectra for $J'<1$ indicates that 
$J_{2}^{c2}$ move closer to $J_{2}^{m}$ as $J'$ decreases with
$J_{2}^{c2}\approx 0.5J'$ for $J'\rightarrow 0$.
As at  $J_2=0$, for Heisenberg chain coupled by $J'$ interactions,
it is likely that, at $J'=0$, the 1D behavior 
of the single Heisenberg chain is unstable to $J_2$ interchain coupling
leading to  N\'eel LRO as soon as $J_2>0$.

Here too, the N\'eel LRO appears only in the limit of  an infinite
number of chain.
The spin-gap remains finite at $M$ fixed for $L\rightarrow\infty$.
The ground-state is a non degenerate singlet.
One has Haldane-like phases.
Instead of the $(\pi,\pi)$ N\'eel LRO one has 
the same phase as the phase of the $M$-leg (unfrustrated) ladder.
In place of the $(\pi,0)$ N\'eel LRO one has a phase,
that 
is probably closely similar to the one of the spin-$M/2$ chain as
found for the 2-leg and 3-leg ladders.
These phases are analogues of 
the 'singlet' and 'Haldane' phases of 2-leg ladder.
Like the 2-leg ladder they have likely an hidden topological LRO.
This and their eventual relation to the chain of integer spins $S>1$
for which string order parameters have been recently studied~\cite{qlsc03}
deserves further investigation.

\section{ Intermediate region }
\begin{figure} 
   \begin{center}
   \resizebox{7cm}{7cm}{\includegraphics{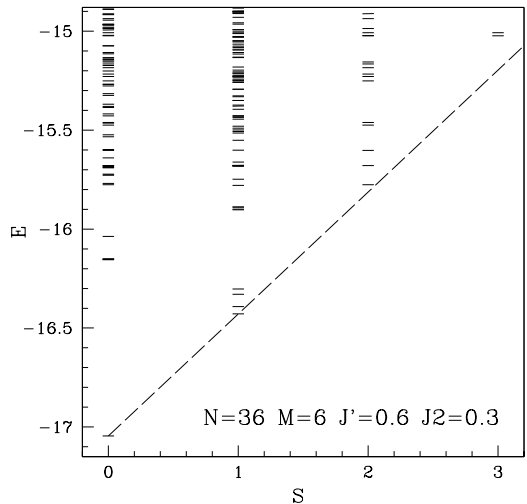}}
   \end{center}
   \caption[99]{ Spectrum of eigenenergies vs total spin $S$
for the $M=6$ sample of $N=36$ spins at $J'=0.6$ and $J_2=0.3$.
These values of $J'$ and $J_2$ correspond to a point
in the intermediate region (slightly) below the line $J_{2}^{m}(J')$
in Fig.~\ref{phase_diag} ($J_{2}^{m}(J'=0.6)\approx 0.32$).
The dashed line, fitted to the lowest $S=1$ and $S=0$ states,
shows that the energy of the lowest states in each spin sector
increase nearly as $\sim S$ at small $S$ values.
        }  \label{go2_36s_6c_j2_0.60_j3_0.30.ps}
\end{figure}

\begin{figure} 
        \begin{center}
        \resizebox{7cm}{7cm}{\includegraphics{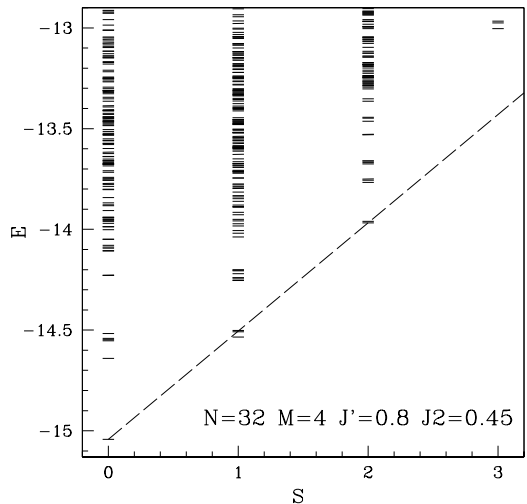}}
        \end{center}
        \caption[99]{ Spectrum
at $J'=0.8$ and $J_2=0.45$
for the $N=32$ sample of $M=4$ chains
        }  \label{spec_32_4c_j2_0.80_j3_0.45}
\end{figure}

Between these two regions of N\'eel behavior, 
  there is  an intermediate region for
the  range of parameters $J_{2}^{c1}(J')\le J_2 \le  J_{2}^{c2}(J')$
  where the spectra indicate an absence of N\'eel LRO
when $L\rightarrow\infty$ and $M\rightarrow\infty$.
Leaving the region of $(\pi,\pi)$ N\'eel LRO by increasing $J_2$ 
at constant $J'$, the evolution of
the energies of the lowest eigenstates in each $S$ sector changes 
to $E(S)-E_0 \sim S$ for small values of $S$
(see Fig.~\ref{go2_36s_6c_j2_0.60_j3_0.30.ps}).
This feature is an indication that the  spin-gap  opens  
for $L\rightarrow\infty$ and $M\rightarrow\infty$~\cite{smlbpwe00}
and one enters a magnetically disordered region. 
At the same time one observes a lowering of some singlet states.
This linear behavior reaches its maximum extension 
for a value of $J_2$ that appears to coincide with $J_{2}^{m}$  
where it extents up to $S=M/2$ if $J' \alt 0.9$ 
as shown in  Fig.~\ref{spec_32_4c_j2_0.80_j3_0.45} for $M=4$
at $J'=0.8$.
Then, beyond $J_{2}^{m}$, the evolution of the lowest eigenenergies
returns fast to $E(S)-E_0 \sim S(S+1)$
in the $(\pi,0)$ N\'eel region which starts at $J_{2}^{c2}$,
close to $J_{2}^{m}$.

\begin{figure} 
        \begin{center}
        \resizebox{7cm}{7cm}{\includegraphics{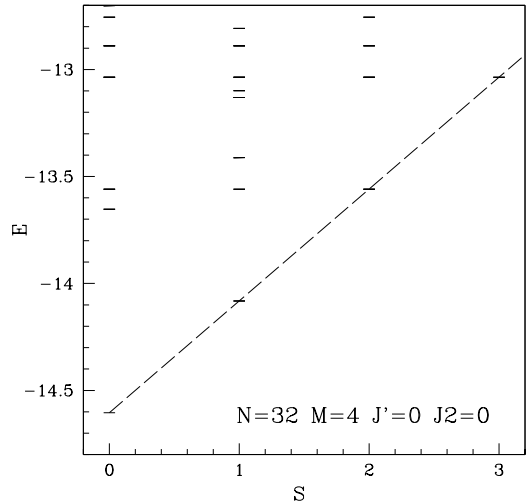}}
        \end{center}
        \caption[99]{ Spectrum of decoupled chains
for the $N=32$ sample of $M=4$ chains
($J'=0$ and $J_2=0$).
Note that the excited levels have a large degeneracy.
        }  \label{spec_32s_4c_indep}
\end{figure}

The quantum numbers of the first excitations in each spin sector
and the signs of the spin-spin correlations remain the same as in the 
$(\pi,\pi)$ N\'eel region  for $J_2 \alt J_{2}^{m}$
and are the same as in $(\pi,0)$ region for $J_2 \agt J_{2}^{m}$. 
The curve $J_{2}^{m}(J')$ is the analogue of the transition
line of the classical model $J_2=0.5J'$.
On this line the classical ground-state consist of decoupled chains.
Then for $J_2\approx J_{2}^{m}(J')$ 
one may notice features, in the spectra 
and the spin-spin correlations (observed for $M=2,4,6$),
that present certain similarities with those
of $M$ decoupled chains  if $J'\alt 0.9$
(and those of $L$ independent chains when $L>M$ and $J'\agt 0.9$).

\begin{figure} 
        \begin{center}
        \resizebox{7cm}{7cm}{\includegraphics{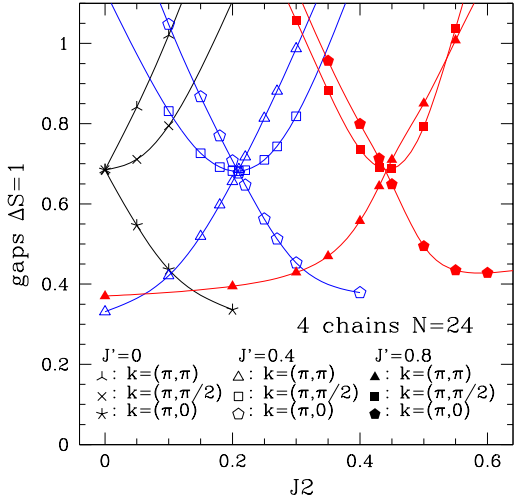}}
        \end{center}
        \caption[99]{ Gaps to the 1th triplet ($\Delta S=1$) vs $J_2$
for $N=24$ samples with $M=4$ (4 chains) and wave vectors $k=(\pi,k_y)$
on the side of the Brillouin zone if $J'=0,0.4,0.8$.
The lines  are guides for the eye.
        }  \label{gaps_s1_n24_m4}
\end{figure}

As indicated above, one has $E(S)-E_0 \sim S$ up to $S=M/2$, 
as for decoupled chains (see Fig.~\ref{spec_32s_4c_indep}),
which is what would result for independent magnetic excitations on the chains.
Moreover, the very lowest eigenstates in each $S$ sector belong to the same IR
as those of a system of decoupled chains.
For instance, the first triplets excitations is a set of $M$ states 
with all wave-vectors of type $(\pi,k_y)$ on the side the Brillouin zone
which become quasi degenerate 
at $J_2\approx J_{2}^{m}$ as shown in Fig.~\ref{gaps_s1_n24_m4} 
for the $M=4$, $N=24$ sample at $J'=0.8$.
Similarly,
first excitations at $S \ge 1$ and $S\le M/2$ have same quantum numbers
(wave vector, characters in point-group symmetries) as
those of decoupled chains which result from the combination
of first triplets excitations.
The energies per spin (see Fig.~\ref{e_j2_0.8_vs_j}) and
the spin gaps at finite size are also 
quasi independent of the number of chains.

In addition the spin excitations appear quasi gapless.
Although the finite-size spin-gaps are maximum for $J_2\approx J_{2}^{m}$,
as shown in Fig.~\ref{gaps_s1_n24_m4}, the analysis of the
evolution of the spin-gap $\Delta^{1}$ of the $M=4$ samples with $1/L$ indicates
that its value at $L\rightarrow\infty$ is minimum for $J_2\approx J_{2}^{m}$
as for the 2-leg ladder~\cite{wang98,hhr01}.
If $J'\le 0.6$ the scaling behavior
of $\Delta^{1}$ is quite similar to the one for the single chain
(its evolution with $1/L$ is quasi linear with a very small
{\it negative} curvature) and  the extrapolated value of $\Delta^{1}$
is $\approx 0$. 
At $J'\agt 0.8$ nevertheless the evolution of $\Delta^{1}$
with $1/L$ turns to have a positive curvature and 
$\Delta^{1}$ extrapolates to a finite (but small) value.
This may be consistent with a spin-gap opening exponentially
with increasing $J'$ along $J_{2}^{m}(J')$ as predicted
by NT at weak interchain coupling along $J_2=0.5J'$.
However the spin-gap on $J_{2}^{m}(J')$
appears to remain very small  up to large
interchain coupling for $M\le 4$ as found for $M=2$~\cite{hhr01}.

\begin{figure} 
        \begin{center}
        \resizebox{7cm}{7cm}{\includegraphics{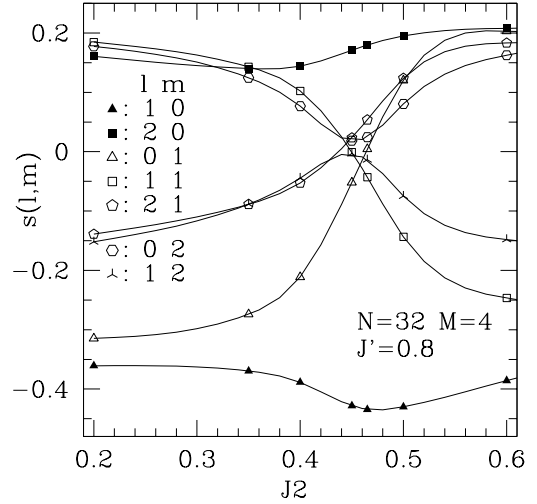}}
        \end{center}
        \caption[99]{ Spin-spin correlations
         $s(l,m)$ (Eq~\ref{eq-SScor}) between spins at distance $(l,m)$
         vs $J_2$ 
         for $N=32$, $M=4$, $J'=0.80$.
The lines  are guides for the eye.
        }  \label{cor_spin_n32_j2_0.8}
\end{figure}

An  apparent decoupling of the chains is also manifest in
the spin-spin correlation between two spins at distance $(l,m)$:
\begin{equation}
s(l,m)= < {\bf S}_{0,0}.{\bf S}_{l,m} > \;.
\label{eq-SScor}
\end{equation}
Values of $s(l,m)$ for the $N=32$ (M=4) sample are displayed 
vs $J_2$ at $J'=0.8$ in Fig.~\ref{cor_spin_n32_j2_0.8}
for  pair of spins on the same ($m=0$) or different ($m=1,2$) chains at short
distances along the chains ($l \le 2$).
Since $s(l,m)$ decrease in magnitude at given $m$ with increasing
distance $l$ along the chains these are the largest values  of $s(l,m)$.
$J_{2}^{m} \approx 0.45$ when $J'=0.8$ and $N=32$.
As shown in Fig.~\ref{cor_spin_n32_j2_0.8} the inter-chain 
spin-spin correlations become very small at $J_2=0.45$:
if $m\ne 0$, $s(l,m)\approx 0$ except between first neighbor spins
which changes of sign at a slightly larger $J_2$ value.

Others features in the spectra  nevertheless testify of 
interactions between the chains when $J_2 \approx J_{2}^{m}(J')$ 
for $M\le 4$ as found for $M=2$.
Starting from situation of decoupled chains at $J'=0$ and
increasing $J'$ along $J_{2}^{m}(J')$ one sees a modification of 
the structure of the eigenlevels above the most lowest eigenstates 
in each $S$ sector. 
Levels which were degenerate for uncoupled chains move away from one-another.
In particular, there is a number
of singlet states which separate from the others and 
drift toward the ground state 
(from an energy $\approx$ twice the spin-gap at $J'=0$ to
an energy $\sim$ to the spin-gap at $J'=0.8$ 
as shown in Fig.~\ref{spec_32_4c_j2_0.80_j3_0.45}  
for the $N=32(M=4\times L=8)$ sample).
This could indicate a behavior at $J_2 \approx J_{2}^{m}$ 
that differs from the one of independent chains
and raises the question wether it corresponds to the RVB state
predicted by NT in the limit of small interchain coupling 
for $J_2=0.5J'$.

We investigate below successivelly the three cases
$J_2<J_{2}^{m}$, $J_2 \approx J_{2}^{m}$ and $J_{2}^{m}>J_2$ 
in more detail.
In order to be with an intermediate region sufficiently wide
and to observe effects of interactions between the chains
on small samples we shall mainly display results 
for a rather large $J'=0.8$ value and discuss their evolution
for smaller and larger values of $J'$.

\subsection{ $J_2< J_{2}^{m}$  }
For $J_2< J_{2}^{m}$ and not to close to $J_{2}^{m}$,
the lowest excited singlet states, noted below $|1>$, $|2>$, $|3>$ are: 
$|1>$ in the trivial IR
(quantum numbers: ${\bf k}=0,R(\pi )=1,\sigma=1$),
$|2>$ [${\bf k}=(\pi,0),R(\pi )=-1,\sigma=1$] and
$|3>$ [${\bf k}=(0,\pi),R(\pi )=-1,\sigma=-1$].

\begin{figure} 
        \begin{center}
        \resizebox{7cm}{7cm}{\includegraphics{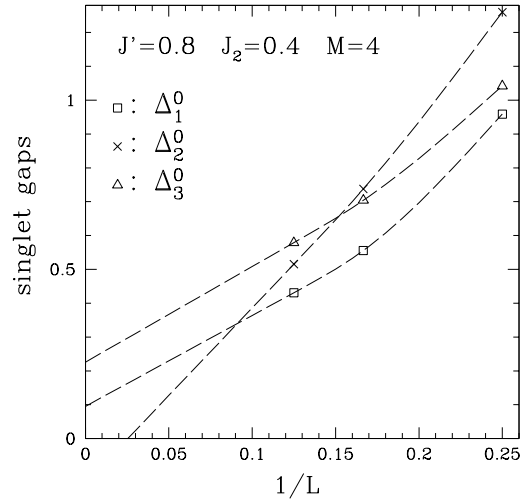}}
        \end{center}
        \caption[99]{ Singlet-gaps $\Delta^{0}_{i}$
at $J'=0.8$ and $J_2=0.4$ vs $1/L$ for
for $N=16,24,32$ samples of $M=4$ chains.
The dashed lines are spline fits to the data for $N \le 32$
followed by linear extrapolation for $N > 32$.
        }  \label{gaps_j2_0.8_j3_0.4_s0}
\end{figure}

Fig.~\ref{gaps_j2_0.8_j3_0.4_s0} shows
the evolution of the singlet-gaps $\Delta^{0}_{i}$ 
from the ground-state $|0>$ to the states $|i>$  
vs $1/L$ at $J'=0.8$ and $J_2=0.4$
for the $M=4$, $N=16,24,32$ samples.
The gaps appear to decrease monotonically.
They evolve with an upward curvature vs $1/L$.
As above we assume that the evolution of the gaps will 
remain similar at larger sizes as found for the 2-leg ladder.
Lower bounds for the gaps are then obtained using
linear extrapolations for $N>32$.
The lower bounds are positive values for
$\Delta^{0}_{1}$ and $\Delta^{0}_{3}$.
This point to finite gaps $\Delta^{0}_{1}$,$\Delta^{0}_{3}$.
On the other hand, the extrapolation 
suggests a possible vanishing of $\Delta^{0}_{2}$ 
for $L\rightarrow\infty$.
A similar analysis of these gaps for the $M=6$, $N=24,36$ samples
also point to the same conclusions for $L\rightarrow\infty$ if $M=6$.

The vanishing of $\Delta^{0}_{2}(L\rightarrow\infty)$ for $M\le 4$
may be questioned since the 2-leg ladder ($M=2$) is known 
to be fully gapped~\cite{hhr01}.
The same analysis indeed shows that $\Delta^{0}_{2}(L\rightarrow\infty)$
is finite and rather large if $M=2$ at $J'=0.8$ and $J_2=0.4$.
From this, it could be argued that $\Delta^{0}_{2}(L\rightarrow\infty)$
decreases with increasing values of $M$ 
but remains finite for $M$ finite.
Our data, limited to small sizes cannot exclude this possibility.
However, $\Delta^{0}_{2}$ most likely vanish on $J_{2}^{m}$
for $M\ge 4$ (see Sec. V B) and then,
the range of values of $J_2$ where $\Delta^{0}_{2}$ appear to vanish,
starts at $J_2\approx 0.35$ and extends beyond $J_{2}^{m}$
till $J_2\approx 0.5$ (see below) if $M=4$.
This interval is large which support a vanishing of $\Delta^{0}_{2}$
on an extended range of values of $J_2$.
An other possibility, is that the value of $J_2$ at given $J'$
beyond which $\Delta^{0}_{2}(L\rightarrow\infty)$ would vanish
vary with $M$ and  drift toward $J_{2}^{m}$ as $M$ decreases. 
ED calculations indeed point out that $\Delta^{0}_{2}(L\rightarrow\infty)$
drops to a minimum at $J_2\approx J_{2}^{m}$ for $M=2$ where
this gap is quasi-vanishing.

On the other hand,
an analysis of the evolution of the spin-gap $\Delta^{1}$ with $1/L$
show that it remain finite if $L\rightarrow\infty$ for $M=4$
and $J_2 \le J_{2}^{m}$.
Besides the fact that $E(S)$ neither  evolve as $E(S)- E_0 \sim S(S+1)$
or as $E(S)-  E_0 \sim S$ indicate that the spin-gap is finite
for $L\rightarrow\infty$ and $M\rightarrow\infty$.

All together, the ED results suggest the occurence of 
a $(\pi,0)$ VBC LRO at $J'=0.8$ for $J_2\agt 0.35$,
breaking translational symmetry in the horizontal direction
with a twice degenerate ground-state,
gapped to others singlet and magnetic excitations, 
possibly as soon as $M=4$  when $M$ is even,
and most likely when $M\rightarrow\infty$.

\begin{figure} 
        \vspace{-2.6cm}
        \begin{center}
        \resizebox{7cm}{7cm}{\includegraphics{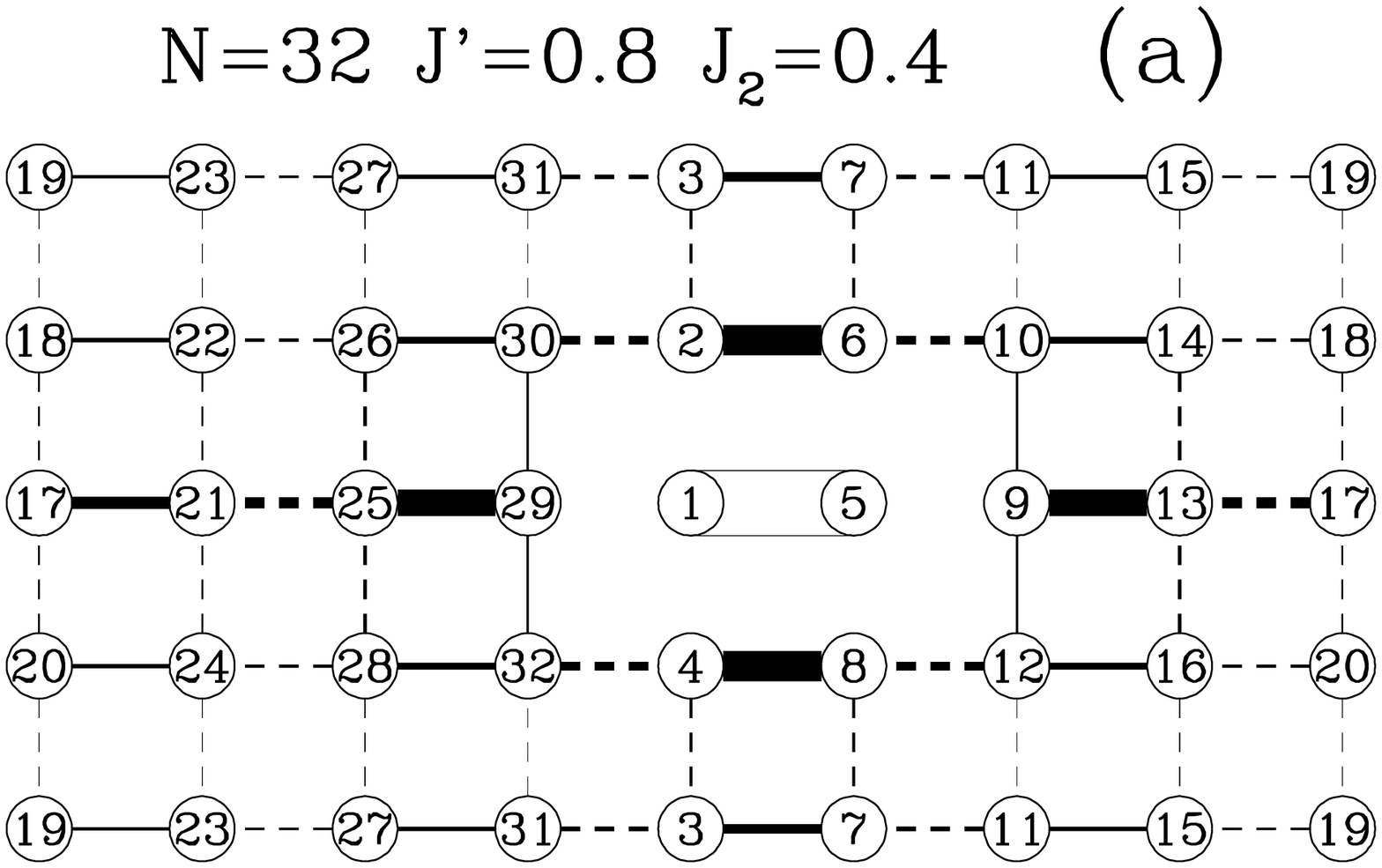}}
        \vspace{-2.8cm}

        \resizebox{7cm}{7cm}{\includegraphics{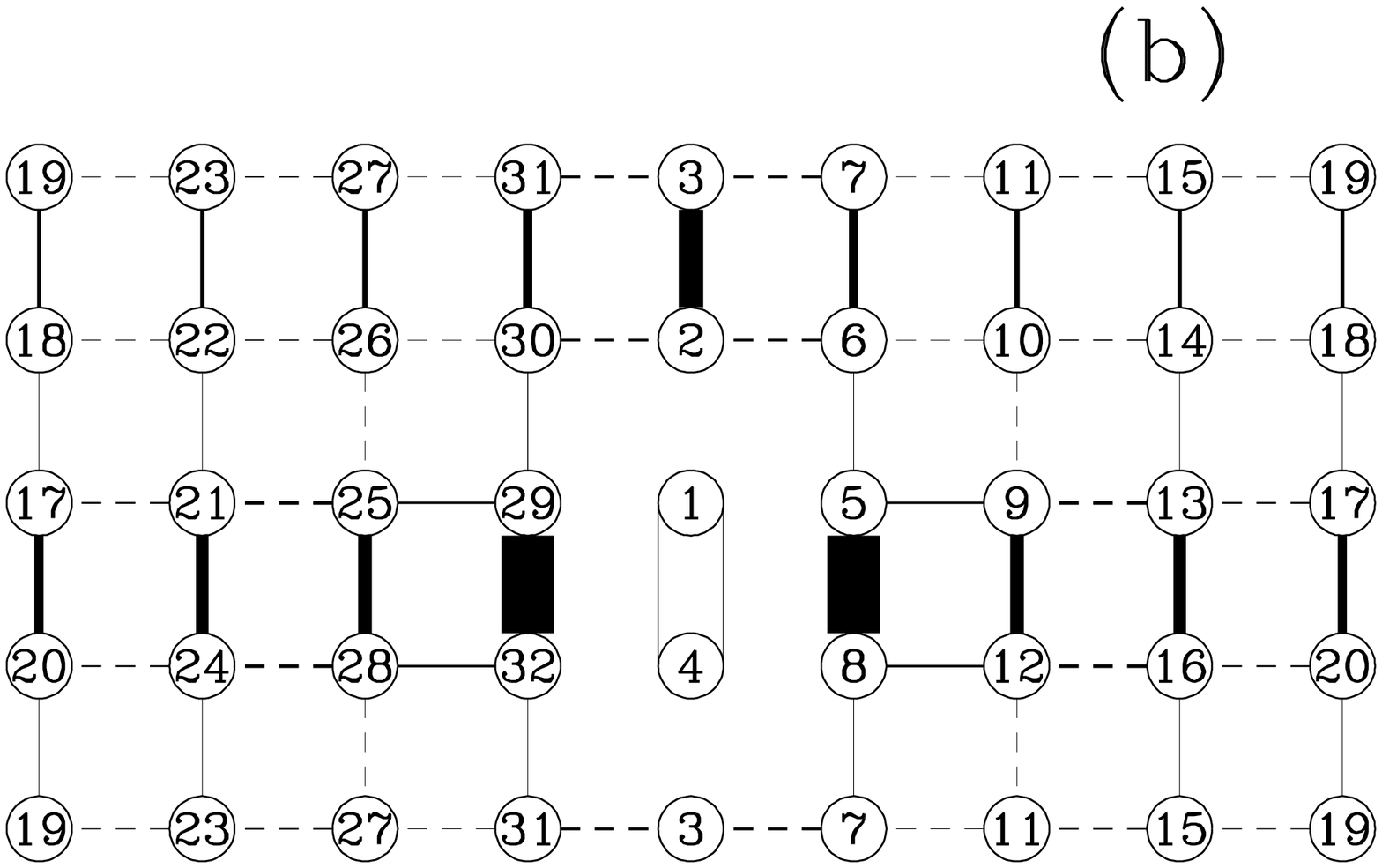}}
        \vspace{-1.8cm}

        \end{center}
        \caption[99]{ Dimer-dimer correlation function
         $D(i,j;k,l)$ (Eq~\ref{eq-DDcor})
        of a dimer on a reference pair of sites $(i,j)$ with
        a dimer on a pair of sites $(k,l)$
                      for $N=32$, $J'=0.80$, $J_2=0.4$.
        The reference bond: $(1,5)$ in $(a)$, $(1,4)$ in $(b)$,
        is represented by a double line.
        A solid (dashed) line means a  positive (negative) value of $D$.
        The thickness of the line is proportional to the magnitude of $D$.
        }  \label{dimer-dimer_j2_0.8_j3_0.4}
\end{figure}

The patterns of dimer-dimer correlations 
\begin{equation}
D(i,j;k,l)= 
< ({\bf S}_{i}.{\bf S}_{j}) ( {\bf S}_{k}.{\bf S}_{l}) >
\,-\, 
<  {\bf S}_{i}.{\bf S}_{j}> < {\bf S}_{k}.{\bf S}_{l} >
\label{eq-DDcor}
\end{equation}
between valence bonds on pairs sites $(i,j)$ and $(k,l)$
are displayed in Fig.~\ref{dimer-dimer_j2_0.8_j3_0.4} 
for the $N=32$ sample at $J'=0.8$ and $J_2=0.4$.
Fig.~\ref{dimer-dimer_j2_0.8_j3_0.4}(a)
is consistent with a columnar $(\pi,0)$ VBC LRO although
the correlations between horizontal dimers on different chains
are rather small.
Fig.~\ref{dimer-dimer_j2_0.8_j3_0.4}(b) shows nonetheless
correlations between vertical dimers
that are not smaller than those between horizontal dimers and
that could be consistent with a columnar $(0,\pi)$ VBC LRO.
This may be related to the fact that 
$\Delta^{0}_{2}$ at $L=8$ is still higher than 
the singlet gaps $\Delta^{0}_{1}$,$\Delta^{0}_{3}$.
It is the extrapolation 
of $\Delta^{0}_{1}$,$\Delta^{0}_{3}$ at $L\rightarrow\infty$
to finite values that exclude a $(0,\pi)$ VBC LRO for $J'=0.8$.
In Fig.~\ref{dimer-dimer_j2_0.8_j3_0.4} we see that the 
dimer correlations are strongest between pair of bonds on
adjacent chains and much weaker otherwise.
These features reinforce with increasing $J_2$ and will be maximum at 
$J_2\approx J_{2}^{m}$.
Simultaneously, the extrapolated values 
of $\Delta^{0}_{1}$,$\Delta^{0}_{3}$ at $L\rightarrow\infty$
decrease toward zero.
If $J'\alt 0.9$, the same features for the singlet gaps and the
dimer correlations remain in the intermediate region 
for $J_2 < J_{2}^{m}$.

Interchanging the role of $J$ and $J'$ in the preceeding analysis
points out that a $(0,\pi)$ VBC could occur for finite $L$ if 
$M\rightarrow\infty$ at sufficiently large $J'$.
On the other hand,
VBC LRO  for finite $M$, in the limit $L\rightarrow\infty$,
may be excluded at large $J'$.
As mentioned in Sec. II, there is a crossover in the range $0.9<J'<1$
to a situation where the systems, for an aspect ratio $M/L<1$,
may be best viewed as an array of $L$ chains in the vertical
directions. 
For larger $J'$ the evolutions of the above singlet gaps 
with $1/L$ show that $\Delta^{0}_{2}$ opens and
all gaps are finite at $L\rightarrow\infty$ for $M=4,6$.
The $(\pi,0)$ VBC LRO then disappear.
In particular, there is no VBC LRO for $M$ finite if $J'=J=1$.
The gaps
nevertheless decrease at fixed $L$ for increasing $M$ and
the value of $J'$  where the crossover occurs for $M/L<1$ 
shifts with increasing values of $M$ to $J'=J$ for $M/L\rightarrow 1$.
Thus  $(\pi,0)$ VBC LRO may extend to $J'\rightarrow J$ 
if $L,M\rightarrow\infty$ where it would be degenerate 
with $(0,\pi)$ VBC LRO. 

However an  analysis of the evolution of the gaps 
$\Delta^{0}_{1}$, $\Delta^{0}_{2}$, $\Delta^{0}_{3}$
with the size of the samples one may study by the ED method
do not allow to conclude that these gaps vanish at $J'=J=1$
for $N\rightarrow\infty$, as required for the  
four-fold degenerate columnar order corresponding to 
degenerate $(\pi,0)$ and $(0,\pi)$ orders.

It could be possible that one has $(\pi,0)$ LRO for $J'<1$
and $(0,\pi)$ LRO for $J'>1$
but that VBC LRO vanishes on the line $J'=J$.
However, the transition out of $(\pi,0)$ VBC phase might be first order:
the gap $\Delta^{0}_{2}$ appears to jump to a large value 
going through the transition by increasing $J'$.
The order parameter would then be finite on the transition line 
and, if this transition line extends to the line $J'=J$ for 
$L=M\rightarrow\infty$, it would imply that the  
four-fold columnar order occurs on the line $J'=J$ 
where the $J-J'-J_2$ model reduces to the $J_1-J_2$ model.
A picture of the dimer correlations similar to
the one observed above for at $J'=0.8$ and $J_2=0.4$
with strongly correlated pair of chains
has been found in dimer series expansions approaches
for the $J_1-J_2$ model when $0.4\alt J_2\alt 0.5$ ~\cite{swho99}.
The pattern of dimer correlations in one of the four-fold degenerate
columnar states  would then be reminiscent of the one off the line $J'=J$.

The locations of the end of the $(\pi,\pi)$ N\'eel phase 
(for $M\rightarrow\infty$)
and of the beginning of the $(\pi,0)$ VBC phase 
may differs~\cite{sow02}.
They are difficult to estimate  accuratelly 
from ED calculations with the present sizes but
are probably very close or coincident as conjectured for
the $J_1-J_2$ model~\cite{sow02}.

\subsection{ $J_2 \approx J_{2}^{m}$  }
$J_{2}^{m}$ is $\approx 0.45$ at $J'=0.8$. 
At $J'=0.8$ and $J_2=0.45$, the evolution of the gaps to the first triplet 
for $M=4$ with $1/L$ (not shown) indicates that the spin-gap remains finite 
$L\rightarrow\infty$ although it  has become much smaller than
at $J_2 < J_{2}^{m}$.
The new feature is the appearence of
other singlet states in the low part of the spectrum
which, at first sight,  appear to form a set of states 
separated from the singlet continuum
(see Fig.~\ref{spec_32_4c_j2_0.80_j3_0.45}).
This set includes the lowest states with wave vector 
${\bf k}=(0,k_y)$ [$k_y=2\pi m/M$ $m\in [M/2-1,...,M/2]$) 
even under $\sigma$: the two states noted 
$|4>$ [${\bf k}=(0,\pm\pi/2)$, $\sigma=1$] if $M=4$
or the states [${\bf k}=(0,\pm\pi/3)$, $\sigma=1$],
[${\bf k}=(0,\pm 2\pi/3)$, $\sigma=1$] if $M=6$,
then noted $|4>$ and $|4'>$.
If $M=4$, there also the 2th excited state in the trivial IR,
 noted $|5>$, just above the states $|1>$, $|2>$, $|3>$, $|4>$ 
whereas if $M=6$, there are also the lowest states
[${\bf k}=(\pi,\pm\pi/3)$, $\sigma=-1$], then noted $|6>$.
Next if $M=4$, one has just below the singlet continuum
a state, noted $|6>$, with  [${\bf k}=(\pi,\pm\pi/2),\sigma=-1$],
whereas if $M=6$, one finds, adjacent to the singlet continuum
the 2th excited state in the trivial IR (noted $|5>$) followed
by states with wave vectors ${\bf k}=(0,k_y)$.

This low energy spectrum is different
from the one for independent chains to which it may be compared.
In the case of decoupled chains the lowest singlets above
the ground-state (see Fig.~\ref{spec_32s_4c_indep})
is the degenerate set ${\cal S}_1$ of $M$ states 
with wave vectors ${\bf k}=(\pi,k_y)$
($k_y=2\pi m/M$ $m\in [M/2-1,...,M/2]$)
which consists of $|2>$, $|6>$ and the state
[${\bf k}=(\pi,\pi),R(\pi )=-1,\sigma=1$], noted $|8>$ if $M=4$
whereas it includes $|2>$, $|6>$, $|8>$ and the state
[${\bf k}=(\pi,\pm 2\pi/3)$, $\sigma=-1$] noted $|8'>$ if $M=6$.
These states correspond to $S=0$ lowest excitations along individual chains 
(which wave vector differs from the one of the ground-state by
$\Delta k=\pi$).
Just above this set, one has the set ${\cal S}_2$
of $C^{2}_{M}$ degenerate states with wave vector ${\bf k}=(0,k_y)$ 
which consists of $|1>$, $|3>$, $|4>$, $|5>$ and the state
[${\bf k}=(0,\pi),R(\pi )=1,\sigma=1$], noted $|7>$ if $M=4$ (6 states),
whereas if $M=6$, it includes $|1>$, $|3>$ $|4>$, $|4'>$, $|5>$, $|8>$,
a 3th state in the trivial IR, a second state in the same IR as $|4>$ and
two states in the same IR as $|4'>$ (15 states).
These states correspond to the combination of two $S=1$ lowest excitations 
on different chains in a total singlet state
(the lowest $S=1$ excitations of the single chain has a wave vector
which differs from the one of the ground-state by $\Delta k=\pi$).
Third, on has a set ${\cal S}_3$ of $M$ states
with wave vectors ${\bf k}=(\pi,k_y)$.
These states are:
[${\bf k}=(\pi,0),R(\pi )=-1,\sigma=-1$],
[${\bf k}=(\pi,\pi/2),\sigma=1$],
[${\bf k}=(\pi,\pi),R(\pi )=-1,\sigma=-1$] if $M=4$.
They correspond to the combination of three lowest $S=1$ excitations
on different chains.
Above ${\cal S}_3$ appear sets of states with wave vectors inside the
Brilloun zone which involve higher excitations
on individual chains at $\Delta k\ne \pi$ 
and appear to form a continuum.
At small values of $L$, the sets of states ${\cal S}_1$, ${\cal S}_2$ 
remain well separated  from this continuum which starts at ${\cal S}_3$.
Yet, in the limit $L\rightarrow\infty$ the gaps to the states
of ${\cal S}_1$, ${\cal S}_2$, ${\cal S}_3$,
vanishes as the gaps to the lowest $S=0,1$ excited states
of a single chain.
The excited states form a gapless continuum adjacent to the ground-state.

\begin{figure} 
        \begin{center}
        \resizebox{7cm}{7cm}{\includegraphics{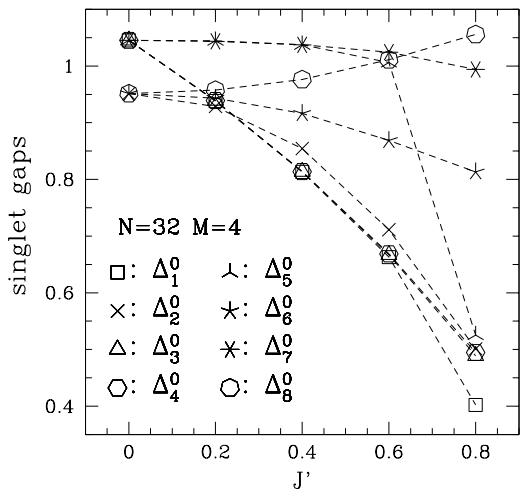}}
        \end{center}
        \caption[99]{ Singlet-gaps $\Delta^{0}_{i}$ (see text)
of the $N=32$ $M=4$ sample vs $J'$  
for values of $J_2\approx J_{2}^{m}(J')$.
The dashed lines are guide to the eyes.
        }  \label{gaps_s0_line}
\end{figure}

On the line of maximum frustration,
the degeneracy in the sets ${\cal S}_1$ and ${\cal S}_2$ 
appears to be lifted by interchain couplings.
Fig.~\ref{gaps_s0_line}  shows the variation of the gaps $\Delta^{0}_{i}$ 
for the $N=32$ sample vs $J'$ for values of $J_2\approx J_{2}^{m}(J')$.
The three states $|2>$, $|6>$, $|8>$ of ${\cal S}_1$ splits.
The gaps $\Delta^{0}_{2}$, $\Delta^{0}_{6}$ decrease 
with increasing $J'$, while $\Delta^{0}_{8}$ increases.
The set ${\cal S}_2$ separate in two groups. 
The first group consists of the states $|1>$, $|3>$, $|4>$
which remain very close in energy up to $J'=0.6$.
Their gaps decrease with increasing $J'$.
The second  group includes the states $|5>$ and  $|7>$ which remain at 
approximativelly the same energy up to $J'\sim 0.6$. 
Beyond this value, $|5>$
decreases rather abruptly to an energy  close to the first group,
a drop which may be attributed to the proximity of the range  
of values of $0.9<J'<1$ where the crossover to the large $J'$
behavior occurs  (for $J'=0.9$ additional singlet states
which evolve from the lowest singlet states of $L$ 
independent chains will begin to appear in the low part 
the spectrum if $N=32$).
A similar splitting also occurs in most of the sets of states
which are above ${\cal S}_3$ 
(the states of ${\cal S}_3$ remain nevertheless quasi-degenerate).
The energy of some of these states and the energy of the states
of ${\cal S}_3$ decrease with increasing $J'$ and become
comparable to the energy of the the states $|5>$, $|7>$, $|8>$ 
for $J'\sim 0.6$.
Yet, up to $J'=0.4$ the evolution vs $1/L$ of the gaps $\Delta^{0}_{i}$
to the states of ${\cal S}_1$ and ${\cal S}_2$
shows that these splitting are much reduced  
for $L\rightarrow\infty$.
All the gaps to the states of ${\cal S}_1$, ${\cal S}_2$ and ${\cal S}_3$
extrapolate to $\approx 0$.
The singlet spectrum thus remains close to the one of 
independent chains.  
The finite-size results give nevertheless a first indication
that the splitting of the states in ${\cal S}_1$, ${\cal S}_2$
will subsist  for $L\rightarrow\infty$.
A differentiation in the extrapolated values
of the gaps becomes  more visible at larger $J'$ values.

\begin{figure} 
        \begin{center}
        \resizebox{7cm}{7cm}{\includegraphics{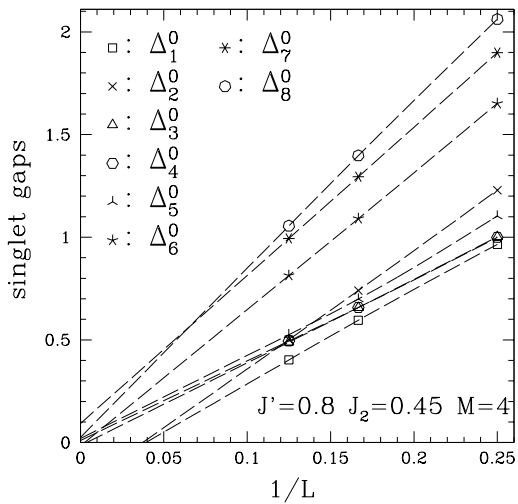}}
        \end{center}
        \caption[99]{ Singlet-gaps $\Delta^{0}_{i}$ (see text)
at $J'=0.8$ and $J_2=0.45$ vs $1/L$ for
for $N=16,24,32$ samples of $M=4$ chains
The dashed lines are spline fit to the data for $N \le 32$
followed by linear extrapolation for $N > 32$.
        }  \label{gaps_j2_0.8_j3_0.45_s0}
\end{figure}

Fig.~\ref{gaps_j2_0.8_j3_0.45_s0} shows
the evolution of the singlet-gaps $\Delta^{0}_{i}$ vs $1/L$ for
the $M=4$ samples at $J'=0.8$ and $J_2=0.45$.
The evolutions of the $\Delta^{0}_{i}$ still remain
quasi-linear with $1/L$ 
if  $L \le 8$ ($N \le 32$) on $J_{2}^{m}(J')$ up to $J'=0.8$.
A curvature is hardly visible.
But, since a linear extrapolation for $N>32$ 
would leads to negative values for 
$\Delta^{0}_{1}$, $\Delta^{0}_{2}$, $\Delta^{0}_{3}$, $\Delta^{0}_{6}$,
an upward curvature may be infered for these gaps. 
This affects the precision of the extrapolations  for $L\rightarrow\infty$
but one may conjecture,  in view of Fig.~\ref{gaps_j2_0.8_j3_0.45_s0},
that not only $\Delta^{0}_{2}$ still likely vanishes as 
at $J_2=0.4$, but also $\Delta^{0}_{1}$, 
whereas the other $\Delta^{0}_{i}$ are very small
and some may vanishes. 
This could be the case of $\Delta^{0}_{3}$, $\Delta^{0}_{6}$
which extrapolate to negative values
and probably $\Delta^{0}_{4}$ which follow closely $\Delta^{0}_{3}$
for all sizes if $J'\le 0.8$.
The vanishing of $\Delta^{0}_{6}$ is nonetheless uncertain as
the degeneracy of state $|6>$ with the state $|2>$ 
seems to be lifted on the line $J_{2}^{m}(J')$.
But  $\Delta^{0}_{7}$, $\Delta^{0}_{8}$ and probably
$\Delta^{0}_{5}$ which extrapolate above $\Delta^{0}_{4}$
remain finite .
Thus the degeneracies which occcur for decoupled chains
are lifted.
Nonetheless these gaps remain probably very small 
which suggest that they open exponentially like the spin-gap.
The gaps to the states at the bottom of the
singlet continuum are also small.
Those to the states of ${\cal S}_3$ extrapolate to quasi-vanishing values.
The spectrum appears to be gapless  or quasi gapless.

The exact degree of ground-state degeneracy is difficult to acertain,
especially for $M=6$.
The most likely vanishing gaps are $\Delta^{0}_{1}$,
$\Delta^{0}_{2}$, $\Delta^{0}_{3}$, $\Delta^{0}_{4}$, $\Delta^{0}_{6}$.
for $M=4$. 
This would lead to a $2^{M-1}=8$ ground-state degeneracy.
Yet the degeneracy  would be limited to $6$ if $\Delta^{0}_{6}$
remains finite.
Besides the gaps to the lowest states in the  singlet continuum
appear to be quasi-vanishing.
For $M=6$, the extrapolations of the singlet gaps, although even more 
unaccurate than for $M=4$, also indicate 
a large ground-state degeneracy 
and a gapless  continuum of singlet excitations adjacent
to the ground-state.
This also suggests a spectrum that correspond closer to the one
of the RVB state of NT than to the one of independent  chains.

\begin{figure} 
        \vspace{-2.6cm}
        \begin{center}
        \resizebox{7cm}{7cm}{\includegraphics{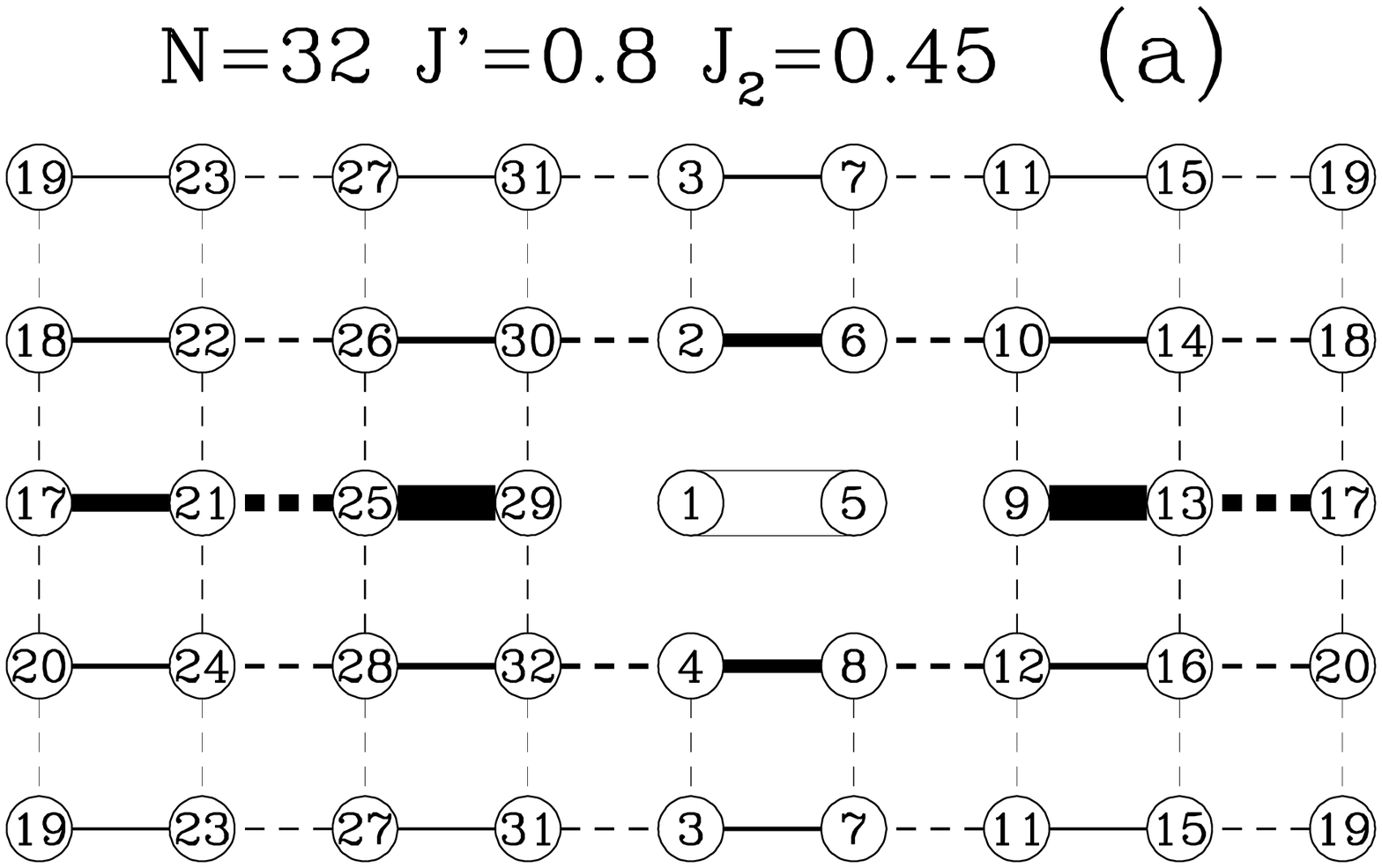}}
        \vspace{-2.8cm}

        \resizebox{7cm}{7cm}{\includegraphics{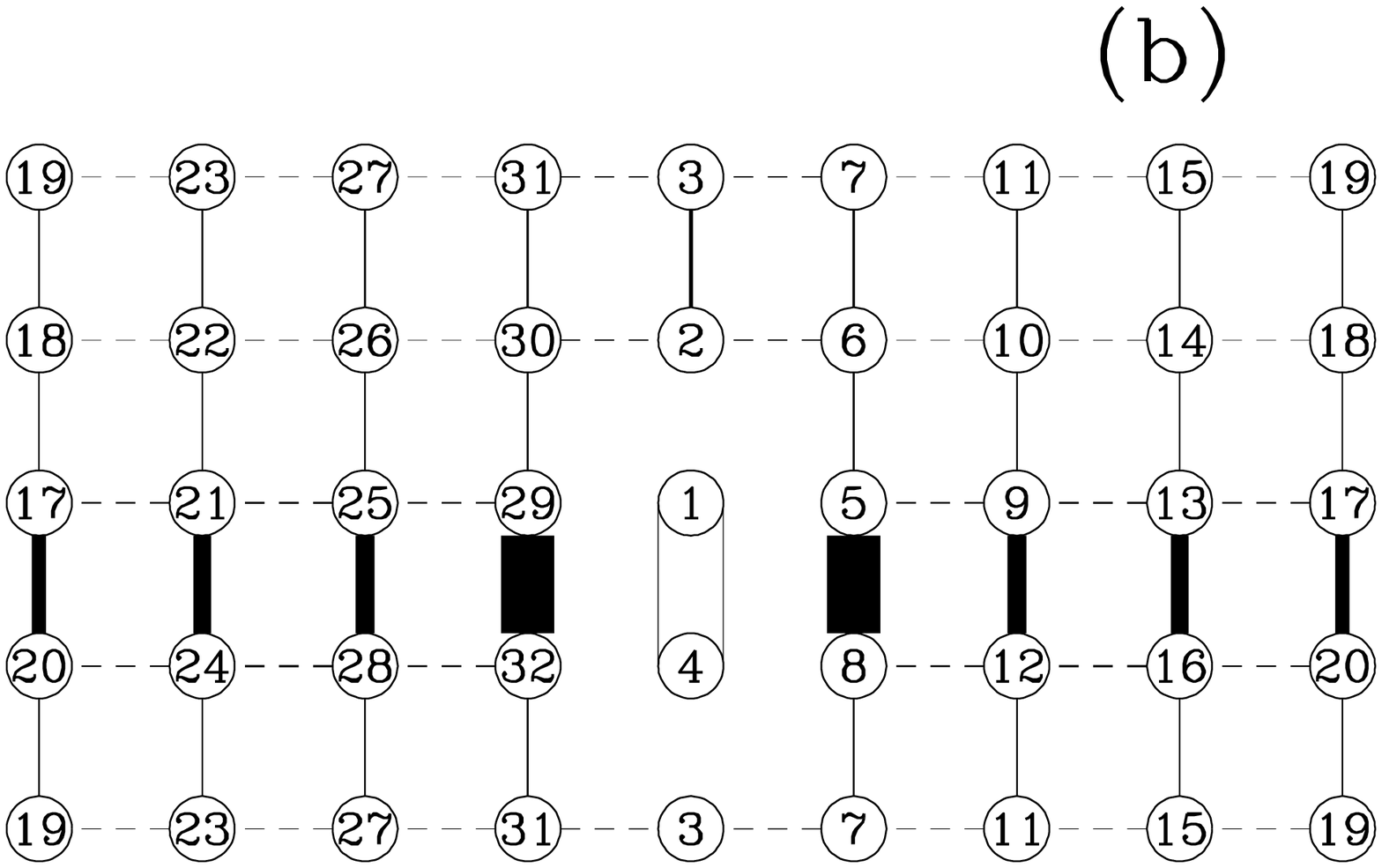}}
        \vspace{-1.8cm}

        \end{center}
        \caption[99]{ Dimer-dimer correlation function
                    (see Fig~\ref{dimer-dimer_j2_0.8_j3_0.4})
                      for $N=32$, $J'=0.80$ $J_2=0.45$.
        }  \label{dimer-dimer_j2_0.8_j3_0.45}
\end{figure}

This degeneracy of  additional singlet states 
is accompagned with a modification of the pattern of 
the dimer-dimer correlations $D(i,j;k,l)$ 
as shown for $N=32$ in Fig.~\ref{dimer-dimer_j2_0.8_j3_0.45}.
Dimer-dimer correlations $D(1,5;k,l)$ 
between a reference horizontal pair $(1,5)$ and a pair $(k,l)$  
on the same chain remain strong but 
have now become extremely small otherwise,
much weaker than for $J_2=0.4$,
as shown in Fig.~\ref{dimer-dimer_j2_0.8_j3_0.45}(a).
As shown in Fig.~\ref{dimer-dimer_j2_0.8_j3_0.45}(b),
this is also the case for
dimer-dimer correlations $D(1,4;k,l)$ between a vertical dimer on a pair
of chains and vertical dimer on a different pair of chains,
whereas $D(1,4;k,l)$ remain of  significant magnitude for pairs of dimers 
joigning the same chains.
Fig.~\ref{dimer-dimer_j2_0.8_j3_0.45}(a)(b) give a picture of
strongly correlated pairs of adjacent chains, 
decorrelated  from one pair to the next with
dimer long range correlations  along individual chains and
along the rungs of pairs of adjacent chains.
This pairing is reminiscent of the RVB state of NT.

The ED  results  thus indicate a behavior close to
the one of independent chains on $J_{2}^{m}(J')$ up to very large
interchain coupling but which  differs.
Instead, they suggest that the RVB behavior predicted by NT 
at weak interchain coupling may  be realized and also 
extends at large interchain coupling, perhaps up to $J'=0.8$,
if $J_2\approx J_{2}^{m}(J')$.

The extension of the RVB state on $J_{2}^{m}(J')$ would however
have an upper limit.
For $J'\agt 0.9$ the evolution of the singlet gaps 
vs $1/L$ show that they are clearly finite 
if $L\rightarrow\infty$ for $M=4$.
This would imply a transition on $J_{2}^{m}(J')$.
The transition might be discontinouus as the gaps 
seems to open abruptly.
An alternative possibility  could be that the RVB state is 
only approximativelly realized on $J_{2}^{m}(J')$.
The degeneracy of the RVB state would be progressivelly
lifted with increasing $J'$ but slower than the 
degeneracy corresponding to the independent chain behavior.

The ED results on the $M=L$ samples of $N=16,36$ spins at $J'=J$
indicate that the behavior of the $J_1-J_2$ model for $J_{2}^{m}$ 
differs from the one  at $J'<J$.
But several features reminiscent of this state subsists.
The lowest triplet excitations have only a small dispersion
on the boundary of the Brillouin  zone.
The singlet spectrum at finite size is rather similar to the one
for $J'<J$:
one finds low lying states with wave vector on the side of the 
Brillouin zone in addition to $|1>$, $|2>$, $|3>$
(degenerate with $|2>$) which would allow to form
a four-fold degenerate columnar VBC.
The $J_1-J_2$ model for $J_{2}^{m}$, if it displays VBC order,
is nonetheless close to a state with a spin-liquid behavior
at this point of the phase diagram.
 
\subsection{ $J_2 > J_{2}^{m}$  }
\begin{figure} 
        \begin{center}
        \resizebox{7cm}{7cm}{\includegraphics{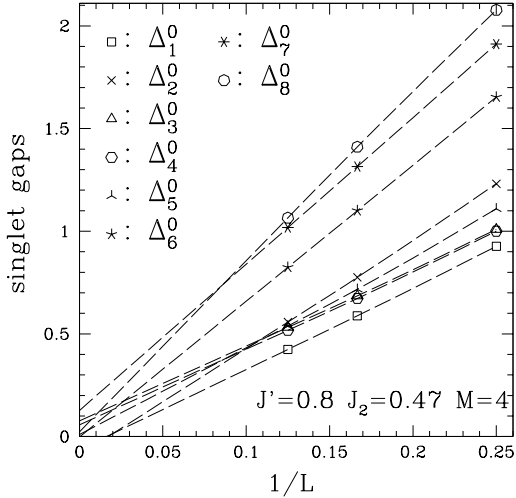}}
        \end{center}
        \caption[99]{ Same as Fig.~\ref{gaps_j2_0.8_j3_0.45_s0}
at $J'=0.8$ and $J_2=0.47$.
        }  \label{gaps_j2_0.8_j3_0.47_s0}
\end{figure}

The behavior predicted by NT is also most likely limited to 
the  line $J_{2}^{m}(J')$.
Increasing $J_2$ at fixed $J'$
one finds a narrow region with a different
behavior before reaching the region where the spectra display
the features of the $(\pi,0)$ N\'eel phase.
The spin-spin correlations $s(l,m)$ (Eq.~\ref{eq-SScor})
at $J'\le 0.8$ have now the same signs as in the
$(\pi,0)$ N\'eel phase: $s(l,m)\sim -1^l$, 
alternating in sign in the horizontal direction
and being ferromagnetic along vertical lines.
Fig.~\ref{gaps_j2_0.8_j3_0.47_s0} shows
the evolution of the singlet gaps at $J'=0.8, J_2=0.47$ for $M=4$ 
which indicate that singlet gaps have opened  except
$\Delta^{0}_{1}$ and  $\Delta^{0}_{2}$ which may still vanish.
This excludes an eight-fold degeneracy of the ground-state.
But the exact degeneracy of the ground-state 
is difficult to acertain. 
The region is very narrow (all singlet gaps are clearly finite
at $J_2=0.5$ when $J'=0.8$), much narrow
than the region between the
the $(\pi,\pi)$ N\'eel phase and  the line $J_2=J_{2}^{m}(J')$.
Due to the proximity of the line $J_2=J_{2}^{m}(J')$ where 
$\Delta^{0}_{1}$, $\Delta^{0}_{2}$ would vanish 
it is difficult to conclude whether  both 
$\Delta^{0}_{1}$ and  $\Delta^{0}_{2}$ really vanishes.

\begin{figure} [h] 
        \vspace{-2.6cm}
        \begin{center}
        \resizebox{7cm}{7cm}{\includegraphics{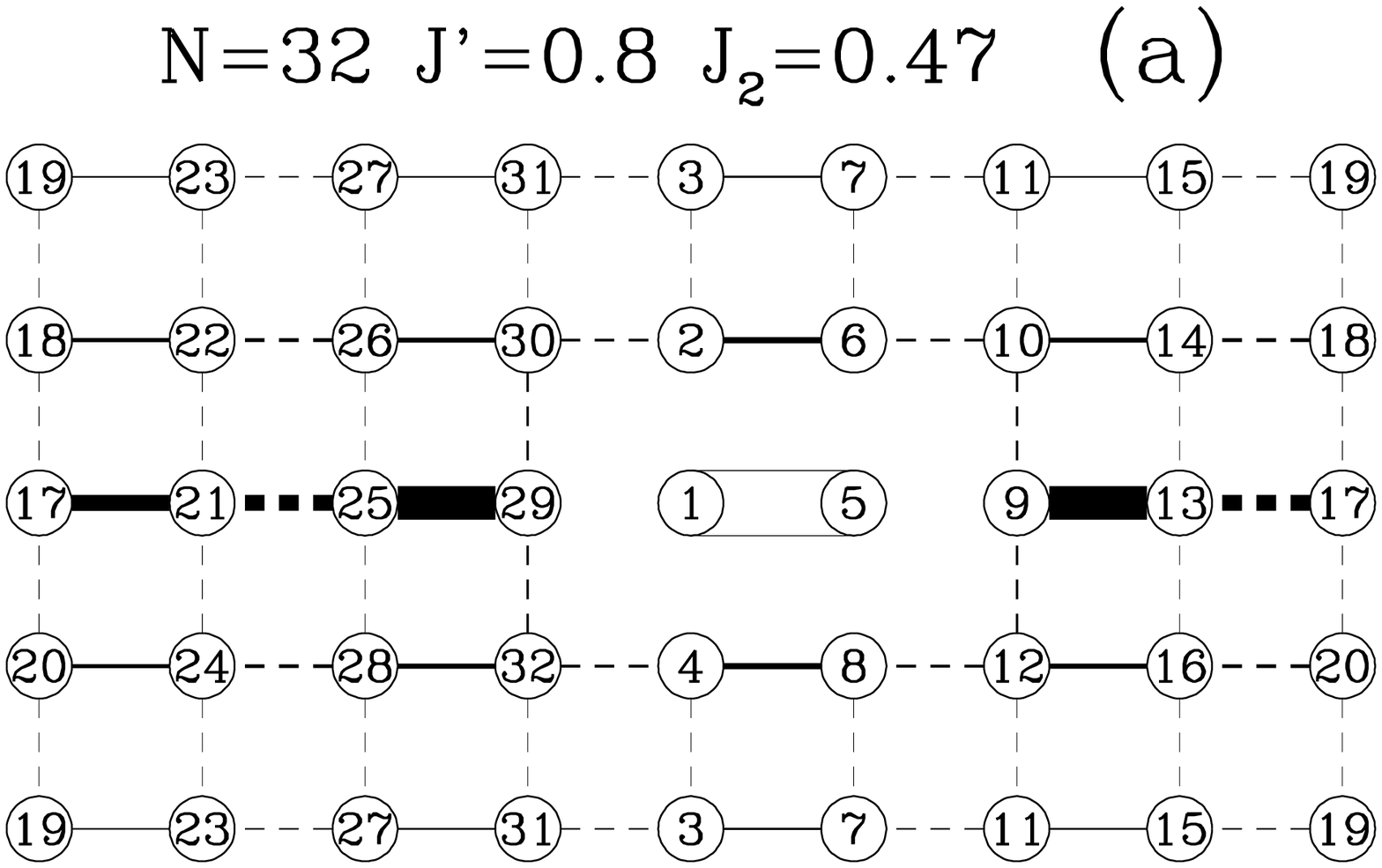}}
        \vspace{-2.8cm}

        \resizebox{7cm}{7cm}{\includegraphics{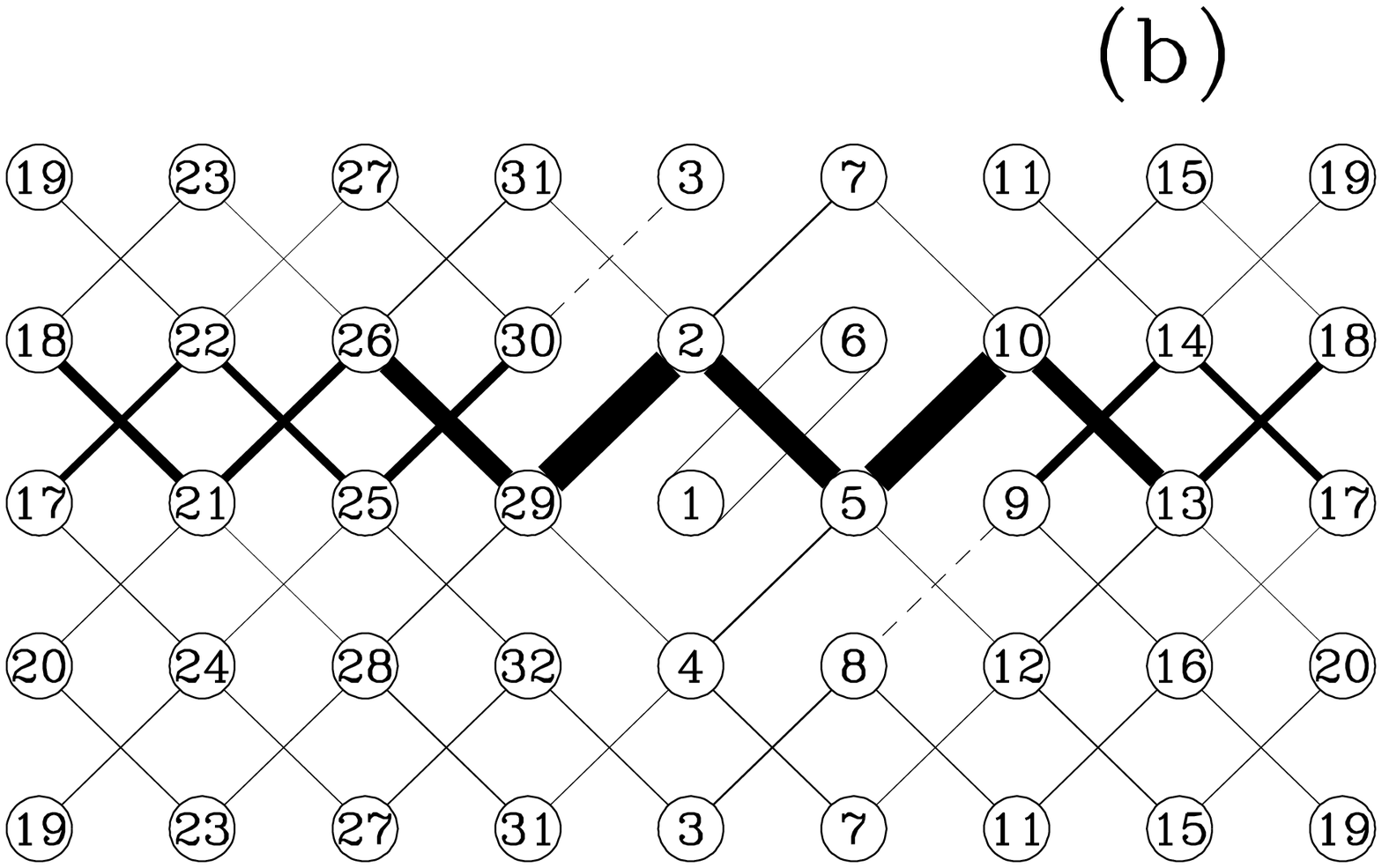}}
        \vspace{-2.8cm}

        \resizebox{7cm}{7cm}{\includegraphics{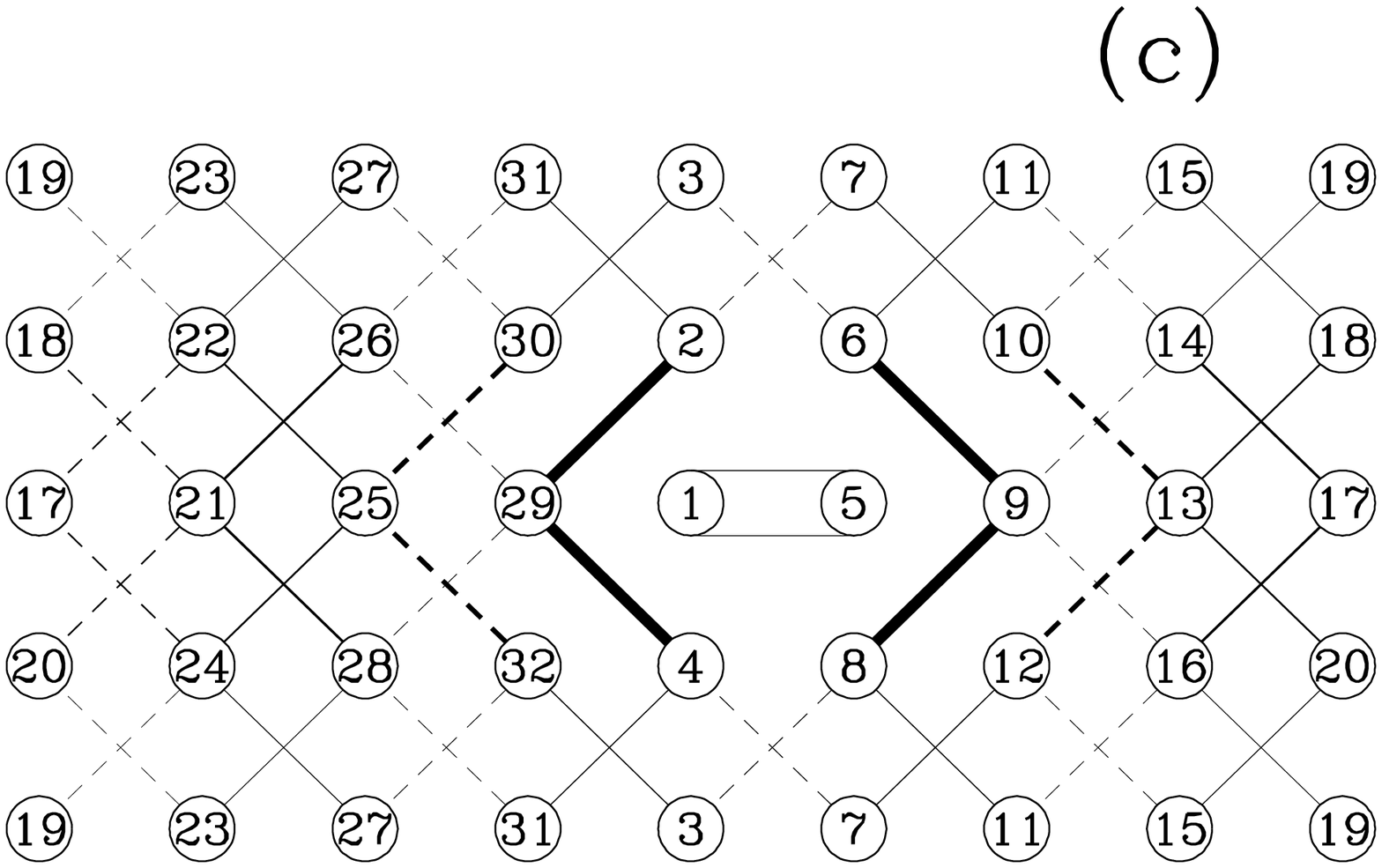}}
        \vspace{-1.8cm}

        \end{center}
        \caption[99]{ Dimer-dimer correlation functions
                    (see Fig~\ref{dimer-dimer_j2_0.8_j3_0.4})
                      for $N=32$, $J'=0.80$ $J_2=0.47$.
        }  \label{dimer-dimer_j2_0.8_j3_0.47}
\end{figure}

The patterns of dimer-dimer correlations are shown 
in Fig.~\ref{dimer-dimer_j2_0.8_j3_0.47} for $N=32$ and $M=4$
at $J'=0.8, J_2=0.47$.
The correlations between horizontal dimers in
Fig.~\ref{dimer-dimer_j2_0.8_j3_0.47}(a)
display a similar alternation  of sign as in
Fig.~\ref{dimer-dimer_j2_0.8_j3_0.4}(a) and
could be compatible with a columnar $(\pi,0)$ VBC LRO,
although they have a small magnitude if the bonds are on 
different chains.

As the shortest pair of spins with the next largest AF spin-spin
correlations are on the diagonals of a square, whereas
the vertical nearest neighbor spins  are ferromagnetically
correlated, we have displayed the correlations between  diagonal bonds
(Fig.~\ref{dimer-dimer_j2_0.8_j3_0.47}(b))
and between  diagonal and  horizontal bonds
(Fig.~\ref{dimer-dimer_j2_0.8_j3_0.47}(c)).

As shown in Fig.~\ref{dimer-dimer_j2_0.8_j3_0.47}(b),
the correlations between diagonal bonds $D(1,6;k,l)$ 
do not show a clear modulation in their magnitudes
neither in the horizontal, the  vertical or the diagonals directions.
$D(1,6;k,l)$ is small
except for bonds linking two adjacent chains.
There subsists a tendency of pairs of adjacent chains
to associate as seen in Fig.~\ref{dimer-dimer_j2_0.8_j3_0.45}(b).
The correlations are then rather similar to those in the 
'Haldane phase' of the two-leg ladder.
There is no visible tendency of the diagonal bonds to order.

As shown in Fig.~\ref{dimer-dimer_j2_0.8_j3_0.47}(c),
the correlations between a horizontal bond and a diagonal bond
in the same column are negative and
there is a sign alternation of the correlations between
an horizontal bond and diagonal bonds in the horizontal 
direction similar to alternation in Fig.~\ref{dimer-dimer_j2_0.8_j3_0.47}(a),
compatible with a breaking of translational symmetry in the 
horizontal direction.

A vanishing of $\Delta^{0}_{1}$, $\Delta^{0}_{2}$ 
would lead to a three-fold ground-state degeneracy,
but the pattern of dimer-dimer correlations do not indicate 
a VBC LRO with a three-fold degeneracy  of states $|0>$, $|1>$, $|2>$.

On the other hand, a columnar VBC LRO with plaquette modulation,
which would lead to an additional translational symmetry breaking in 
the vertical direction, appears also unlikely for $J' \le 0.8$ 
in view of the pattern of 
dimer-dimer correlations and the fact that the gap $\Delta^{0}_{3}$
to the singlet state $|3>$ with wave-vector $(0,\pi)$ is finite.

If there is VBC LRO, it is most probably a $(\pi,0)$ like VBC LRO.
It will be associated with a vanishing of $\Delta^{0}_{2}$ whereas
$\Delta^{0}_{1}$ will be finite. 
The small but noticable upward curvature in the evolution of $\Delta^{0}_{1}$
in Fig.~\ref{gaps_j2_0.8_j3_0.47_s0} supports indeed that 
$\Delta^{0}_{1}$ might be finite for $L\rightarrow \infty$ in 
this region for $J'=0.8$. 
As for $J_2<J_{2}^{m}$, the $(\pi,0)$ VBC LRO  disappears for
for $J'<1$ if $M$ is finite but could extend for $J'\rightarrow 1$
if $M=L\rightarrow\infty$.

The occurence of a $(\pi,0)$ VBC at $J_2 < J_{2}^{m}(J')$  
would mean that there is the same kind of VBC LRO
on both sides of $J_{2}^{m}(J')$ for $M \ge 4$.
The state for $J_2 > J_{2}^{m}(J') $ could however
differs from the state for $J_2 < J_{2}^{m}(J')$ 
by some non-local order parameter similar to those of
the 'singlet state' and the 'Haldane state' 
of the two-leg ladder as conjectured at the end of Sec. IV
for the phases at finite $M>2$ on both sides of the 
intermediate region.

\section{Summary}
Exact diagonalization calculations provide support 
that the RVB behavior predicted  by Nersesyan and Tsvelik
in  the $J-J'-J_2$ model  at $J_2=0.5J'\ll J$ may be realized 
and extends along a curve coincident with the line of maximum frustration 
$J_{2}^{m}(J')$ where it subsists up to large interchain coupling for
an even number $M$ of chains of lenght $L\rightarrow\infty$.
The line $J_{2}^{m}(J')$ is the analogue of the 
transition line $J_2=0.5J'$ in the classical model 
on which the chains decouples. The independent chain 
behavior appears destabilized by quantum fluctuations.

The line $J_{2}^{m}(J')$ is located in an intermediate region of the phase
diagram where the present results suggest that the ground-state
displays a columnar $(\pi,0)$ valence bond long range order.
This region of columnar order  will  have a finite width
as soon as $M\ge 4$  and extends on both sides of $J_{2}^{m}(J')$
up to a value of $J'$ close but smaller than $J$ for finite $M$,
that increase with $M$ and approach $J'=J$ for $M=L\rightarrow\infty$.
It separates two phases that are fully gapped Haldane like phases 
for  finite $M$ and evolves respectivelly to the
$(\pi,\pi)$ and $(\pi,0)$ N\'eel phases for $M\rightarrow\infty$.
The Haldane like phases have probably a topological order.
Its nature for $M>2$ deserves further investigation.

The occurence of a columnar order for $J'$ smaller than $J$,
next to the N\'eel phase, 
agrees with the scenario deduced from large-$N$ approaches
that valence bond crystal order is expected
after the destabilization of a collinear AF N\'eel phase.
It is possible that
the columnar order may extend for  $J'\rightarrow J$ 
if  $M=L\rightarrow\infty$ and  that
this scenario may be realized also in the $J_1-J_2$ model 
which corresponds to $J'=J$.
The behavior of the $J-J'-J_2$ model in the intermediate region
for $J'\rightarrow J$ requires nonetheless further study
to confirm this hypothesis.

The exact diagonalization method is limited to systems of
small sizes. 
The extrapolations to the infinite lenght limit are affected 
by uncertainities.
Indeed, some one dimensional systems
requires data at very large lenghts to allow a reliable
extrapolation to the infinite lenght limit.
For certain systems there may be a crossover in the scaling
behavior of the gaps at a very large lenght as 
shown for instance in Ref.~\cite{lqxcts03} where
some gaps do not decrease monotonicaly 
with the lenght beyond a certain lenght.
For the present model we have assumed a monotonicaly decreasing
smooth evolution of the gaps with increasing lenght.
The assumption appears reasonable in view of the data
available at large lenghts in the case of two chains
(the 2-leg ladder). 
Yet, a small change in the evolution of the gaps at 
very large lengths is not to be excluded for those gaps
that appear to extrapolate to vanishing or quasi vanishing values. 
The scaling behavior of the gaps to the low lying
excited states as a function of the number of chains
deserves further studies.
In particular, calculations  using an approach
that can deal with very large systems,
like possibly the DMRG method, would be worth 
to firmly establish wether the gaps that have been 
conjectured to extrapolate to zero really vanish
or are just small but finite in the thermodynamic limit
and thus check the exact degeneracy of the ground state
on the line of maximum frustration.

This study was limited to the case of AF couplings.
Yet, the $J-J'-J_2$ model with  AF intrachain coupling ($J>0$), but
interchain couplings ($J',J_2$) that are either or both 
ferromagnetic (FM) is also of interest~\cite{ack-referee}.
Indeed, FM couplings may be relevant to modelize certain
2-leg ladder materials~\cite{johnston-etal-00} and 
might be introduced to describe some $M$-leg compounds.
 
Classically, the ground-state of the $J-J'-J_2$ model remains AF
for $J'>-1$ and $J_2>-0.5$ (if $J=1$). It is the  $(\pi,\pi)$ 
N\'eel state for $J_2<0.5J'$, the  $(\pi,0)$ N\'eel state for $J_2>0.5J'$,
and the $J$ chains stay decoupled for $J_2=0.5J'$.

The spin-1/2, 2-leg ladder (with FM $J'$ and/or $J_2$ ) 
has been studied by bosonization~\cite{kfss00}
and DRMG~\cite{zwc01} calculations. 
These studies showed that the singlet and Haldane phases 
extends for $J',J_2<0$, separated by a transition line 
which is the continuation of the boundary line for $J',J_2>0$.
This transition line is also the line of the points of
maximum frustration $J_{2}^{m}(J')$.
The transition line, however, does not deviate from
the classical boundary line $J_2=0.5J'$ and
the transition appears to be 2nd order for $J',J_2 \le 0$,
instead of 1th order for $J',J_2>0$.
In the bosonization approach, the interchain interaction,
derived at 1th order, proportional to $g=J'+2J_2$ which now turn to be $<0$,
is then irrelevant and does not open a gap.
The system is predicted to be critical and to consist of
independent chains~\cite{philippe}.

A similar change of sign of the interchain interactions
appears in the bosonization approach of NT if $J_2=0.5J'<0$ 
for the 4-leg ladder.
The same critical behavior could occur on the transition 
line  in the 4-leg ladder and perhaps for $M$-leg ladders ($M>2$ even).
Preliminary calculations indicate that the curve  $J_{2}^{m}(J')$ 
also remains on the classical boundary line for $M=4$, where
the model might be critical, perhaps down to $J'=-0.8$.  
The model  may exhibit "sliding Luttinger spin-liquid" behavior
as proposed for the crossed chain model~\cite{ssl02},
at least for not too strong interchain interactions.
On the line $J_2=0.5J'$, the lowest energies $E(S)$ in spin sector $S$
evolves linearly with $S$.
For $M\rightarrow\infty$, 
there may be also an intermediate region 
with a spin-liquid state around this line
in between the N\'eel phases, where $E(S)-E_0\sim S(S+1)$.
This issues deserve further study. 

Acknowledgments: I thank Patrick Azaria, Philippe Lecheminant 
and Claire Lhuillier for fruitful discussions.
I thank the Aspen Center for Physics, where this work was continued,
for hospitality.
Computations were performed at The Centre de Calcul pour la Recherche de
l'Universit\'e Pierre et Marie Curie and at the Institut de
d\'eveloppement des Recherches en Informatique Scientifique du
CNRS under contract 981052.


\begin{thebibliography}{10}

\bibitem{rs90}
N. Read and S. Sachdev, Phys. Rev. B {\bf 42},  4568  (1990).
\bibitem{rs91}
N. Read and S. Sachdev, Phys. Rev. Lett. {\bf 66},  1773  (1991).
\bibitem{sp01}
S. Sachdev and K. Park, Annals of Physics (N.Y.) {\bf 298}, 58 (2002).

\bibitem{fsl01} 
J.-B. Fouet, P. Sindzingre, and C. Lhuillier, 
Eur. Phys. J. B {\bf 20}, 241 (2001).

\bibitem{sfl02} 
P. Sindzingre, J.-B. Fouet, and C. Lhuillier, 
 Phys. Rev. B {\bf 66}, 174424 (2002).

\bibitem{fmsl01}
J.-B. Fouet, M. Mambrini, P. Sindzingre, and C. Lhuillier, 
 Phys. Rev. B {\bf 67}, 054411 (2003).

\bibitem{lws02} 
A. L\"auchli, S. Wessel and  M.Sigrist,
Phys. Rev. B {\bf 66}, 014401  (2002).


\bibitem{sz92}
H. Schulz and T.~A.~L. Ziman,
Eur. Phys. Lett. {\bf 18}, 355 (1992)

\bibitem{szp96}
H. Schulz, T.~A.~L. Ziman and D. Poilblanc,
J. Phys. I (France) {\bf 6}, 675 (1996).

\bibitem{swho99}
R.~R.~P. Singh, Z. Weihong, C.~J. Hamer, and J. Oitmaa,
Phys. Rev. B {\bf 60}, 7278 (1999).


\bibitem{kosw00}
 V.~N. Kotov , Jaan Oitmaa, Oleg Sushkov, Zheng Weihong,
 Phil. Mag. B {\bf 80}, 1483 (2000).

\bibitem{co00}
L. Capriotti and S. Sorella,
Phys. Rev. Lett. {\bf 84}, 3173 (2000).

\bibitem{cls00}
M.~S.~L. du Croo de Jongh, J.~M.~J. van Leeuwen, W. van Saarloos,
Phys. Rev. B {\bf 62}, 14844 (2000).

\bibitem{sow01}
O.~P. Sushkov, J. Oitmaa, and Z. Weihong,
Phys. Rev. B {\bf 63}, 104420 (2001).

\bibitem{cbps01}
 L. Capriotti, F. Becca, A. Parola and S. Sorella,
 Phys. Rev. Lett. {\bf 87}, 097201 (2001).

\bibitem{sow02}
O.~P. Sushkov, J. Oitmaa, and Zheng Weihong,
Phys. Rev. B {\bf 66}, 054401 (2002).

\bibitem{cbps03}
 L. Capriotti, F. Becca, A. Parola and S. Sorella,
Phys. Rev. B {\bf 67}, 212402 (2003).

\bibitem{nersesyan-tsvelik-03}
A.~A. Nersesyan and A.~M. Tsvelik,
 Phys. Rev. B {\bf 67}, 024422 (2003).

\bibitem{smirnov-tsvelik-03}
F.~A. Smirnov and A.~M. Tsvelik, 
 Phys. Rev. B {\bf 68}, 144412 (2003).

\bibitem{bhaseen-tsvelik-03}
M. J. Bhaseen and  A. M. Tsvelik, 
Phys. Rev. B {\bf 68},  094405  (2003).


\bibitem{wko98}
Zheng Weihong, V. Kotov and J. Oitmaa
Phys. Rev. B {\bf 57},  11439 (1998).

\bibitem{wang98}
Xiaoqun Wang, condmat/9803290 (unpublished); 
Mod. Phys. Lett. B {\bf 14}, 32 (2000).

\bibitem{hhr01}
T. Hakobyan, J. H. Hetherington, and M. Roger,
Phys. Rev. B {\bf 63}, 144433 (2001)

\bibitem{aen00}
Dave Allen, F.~H.~L. Essler and A.~A. Nersesyan,
 Phys. Rev. B {\bf 61},  8871 (2000).

\bibitem{wzc02}
Xiaoqun Wang, N. Zhu and C. Chen,
Phys. Rev. B {\bf 66},  172405 (2002).

\bibitem{haldane_83}
F. D. M. Haldane, Phys. Lett. {\bf 93A }, 464 (1983)

\bibitem{wns94}
S. R. White, R. M. Noack, and D. J. Scalapino,
Phys. Rev. Lett. {\bf 73}, 886-889 (1994).

\bibitem{gbw96}
M. Greven, R. J. Birgeneau, U.-J. Wiese,
Phys. Rev. Lett. {\bf 77}, 1865 (1996).

\bibitem{fat96}
B. Frischmuth, B. Ammon, and M. Troyer,
Phys. Rev. B {\bf 54}, R3714 (1996).

\bibitem{sandvik99}
A.~W. Sandvik, Phys. Rev. Lett. {\bf 83}, 3069  (1999).

\bibitem{white96}
S.~R. White, Phys. Rev. B {\bf 53}, 52 (1996).

\bibitem{aln98}
P. Azaria, P. Lecheminant and  A.~A. Nersesyan,
Phys. Rev. B {\bf 58}, R8881 (1998)

\bibitem{moukouri03}
S. Moukouri, cond-mat/0305608 (unpublished).

\bibitem{blp92}
B. Bernu, C. Lhuillier, and L. Pierre, 
Phys. Rev. Lett. {\bf 69}, 2590 (1992).


\bibitem{hn93}
P. Hasenfratz and F. Niedermayer,
Z. Phys. B. Condensed Matter {\bf 92}, 91 (1993).

\bibitem{qlsc03}
Shaojin Qin, Jizhong Lou, Liqun Sun and Changfeng Chen,
Phys. Rev. Lett. {\bf 90}, 067202 (2003).

\bibitem{smlbpwe00}
P. Sindzingre, G. Misguich, C. Lhuillier, B. Bernu, L. Pierre,
Ch. Waldtmann, and H.-U. Everts
Phys. Rev. Lett. 84, 2953 (2000).


\bibitem{lqxcts03}
Jizhong Lou, Shaojin Qin, Tao Xiang, Changfeng Chen, Guang-Shan Tian,
and Zhaobin Su,
Phys. Rev. B {\bf 68}, 045110 (2003).

\bibitem{ack-referee}
This case has been pointed by one the referee.
I thank him/her for this remark.

\bibitem{johnston-etal-00}
D.~C. Johnston, M. Troyer, S. Miyahara, D. Lidsky, K. Ueda, M. Azuma
Z. Hiroi, M. Takano, M. Isobe, Y. Ueda, M.~A. Korotin, V.~I. Anisimov, 
A.~V.  Mahajan, L.~L. Miller,
cond-mat/0001147 (unpublished).

\bibitem{kfss00}
Eugene~H. Kim, G. F\'ath, J. S\'olyom and D.~J. Scalapino,
Phys. Rev. B {\bf 62}, 14965 (2000).

\bibitem{zwc01}
Ningsheng Zhu, Xiaoqun Wang, Changfeng Chen,
Phys. Rev. B {\bf 63}, 012401 (2001).

\bibitem{philippe}
At 2nd order appears an additional contribution to the interchain interaction
that could open a gap.
However, this gap would be likely very small,
probably much smaller than in the case of AF interchain coupling 
and difficult to see numerically.

\bibitem{ssl02}
 O.~A. Starykh, R.~R.~P. Singh, and G.~C. Levine, 
 Phys. Rev. Lett. {\bf 88}, 167203  (2002).


\end{thebibliography}

\end{document}